\documentclass[pdflatex,sn-basic]{sn-jnl}

\usepackage{graphicx}%
\usepackage{multirow}%
\usepackage{amsmath,amssymb,amsfonts}%
\usepackage{amsthm}%
\usepackage{mathrsfs}%
\usepackage[title]{appendix}%
\usepackage{xcolor}%
\usepackage{textcomp}%
\usepackage{manyfoot}%
\usepackage{booktabs}%
\usepackage{algorithm}%
\usepackage{algorithmicx}%
\usepackage{algpseudocode}%
\usepackage{listings}%

\usepackage{aas_macros}

\theoremstyle{thmstyleone}%
\theoremstyle{thmstyletwo}%

\theoremstyle{thmstylethree}%

\begin{document}

\title[Article Title]{The Ubiquity of White-Light Flares on the Sun}


\author[1,2]{\fnm{Shuanghong} \sur{Li}}
\author*[2]{\fnm{Yongliang} \sur{Song}}\email{ylsong@nao.cas.cn}
\author[1]{\fnm{Kun} \sur{Wang}}
\author[2,3]{\fnm{Mei} \sur{Zhang}}
\author[2,3]{\fnm{Xianyong} \sur{Bai}}
\author[2]{\fnm{Xiao} \sur{Yang}}

\affil[1]{School of Physics and Astronomy, China West Normal University, Nanchong 637009, China}
\affil[2]{Key Laboratory of Solar Activity and Space Weather, National Astronomical Observatories, Chinese Academy of Sciences, Beijing 100101, China}
\affil[3]{School of Astronomy and Space Sciences, University of Chinese Academy of Sciences, Beijing 100049, China}


\abstract{
White-light flares (WLFs) have long been considered rare on the Sun, as most magnetic reconnections occur in the corona,  making it difficult to transport sufficient energy downward to heat the very low layers for WL continuum emission. However, it is also argued that WL continuum emission is a common feature of all flares, and the rare detection of them is largely due to the limited sensitivity and resolution of telescopes. Recently, we developed a new method to identify WLFs, which exhibits extraordinary detection capability. Combined with SDO/HMI's over 15 years of continuous photospheric observations, it provides us an unprecedented opportunity to address the long-standing debate about WLF ubiquity. A total of 1951 flares (547 B-class, 1316 C-class, and 88 M-class) from 769 low-activity active regions (ARs; flare number $<$ 10) observed during January 2011 – May 2025 are investigated. Among these, 1012 are identified as WLFs, giving an overall occurrence rate of 51.9\%. The occurrence rates for B-, C-, and M-class flares are 36.9\%, 56.9\%, and 85.2\%, respectively, showing a clear increase with flare class. The occurrence rate is also highly correlated with the AR's sunspot number. For the first time, we find that WL continuum radiation is widely present even among B-class flares. Different WLFs contain varying numbers of WL kernels, which exhibit distinct properties, evolutionary behaviors and different relationships with GOES soft X-ray flux and its time derivative. These results strongly suggest that WLFs are a common phenomenon on the Sun.
}




\maketitle
\section{Introduction}\label{sec1}

Solar flares represent one of the most violent transient energy release phenomena. The released magnetic energy propagates outward in the forms of electromagnetic radiation, energetic particles, and plasma ejections \citep[e.g.][]{Priest2002, Fletcher2011}. These emissions and ejections propagate into interplanetary space, exerting significant influences on planets and other celestial bodies within the solar system, thereby driving space weather effects \citep[e.g.][]{Hanslmeier2007, Lammer2012}. Studies of flares will deepen our understanding of solar magnetic field evolution and energy release processes, and also improve our capability to forecast extreme space weather events and mitigate their impacts on human activities. Among the numerous solar flares, there is a particular class—those that exhibit obvious white-light continuum enhancements, known as white-light flares (WLFs).

The first recorded WLF was independently observed by Carrington and Hodgson in 1859 \citep{Carrington1859, Hodgson1859}. In the past nearly 170 years, WLFs are relatively rare in observations and are regarded as representatives of some extreme conditions of solar flares, posing great challenges to our understanding of atmospheric models and energy transport mechanisms \citep[e.g.][]{Neidig1989, Ding2007}. WLFs are also similar to stellar flares in many respects \citep[e.g.][]{Worden1983, Neidig1989, Kretzschmar2011, Kowalski2017, Kowalski2024, Zhao2024}. The corresponding studies will not only advance our understanding of stellar activity but also shed light on exo-space weather, providing key insights into the potential for extraterrestrial life \citep[e.g.][]{Lammer2012, Airapetian2020}.

Early studies based on ground-based observations suggested that WLFs were very rare, with occurrence rates of only a few events per year \citep{McIntosh1972, Neidig1983}. Before the era of the Solar Dynamics Observatory \citep[SDO;][]{Pesnell2011}, only about 150 flares were conclusively identified as WLFs \citep{Fang2013}. With the observations of Helioseismic and Magnetic Imager \citep[HMI;][]{Scherrer2011} onboard SDO, more WLFs were detected through the Fe {\footnotesize I} 6173 \AA\ continuum \citep[e.g.][]{Kuhar2016, Namekata2017, Song2018c, CastellanosDurn2020, Cai2024}. Recently, China launched the Advanced Space-based Solar Observatory \citep[ASO-S;][]{Gan2023} and Chinese H$\alpha$ Solar Explorer \citep[CHASE;][]{Lichuan2022}. WLFs were also reported using observations in the Balmer continuum (3600 $\pm$ 20 \AA\ ) from the White-light Solar Telescope \citep[WST;][]{Lihui2019} onboard ASO-S and the nearby continua of Fe {\footnotesize I} 6569 \AA\ and H$\alpha$ 6563 \AA\ from CHASE \citep[e.g.][]{Jing2024, Liying2024a}. However, despite the continuous monitoring of flares by multiple satellites, only a few hundred WLFs have been reported so far—a mere fraction of the vast number of solar flares. Notably, most of them are relatively large flares, typically of X- and M-class.

On the other hand, several small C-class WLFs ($<$ C5.0) were observed \citep{Hudson2006, Jess2008, Song2020, Liqiao2024}. It appears to support the idea that WL continuum emission is a universal feature of all solar flares, particularly given that WLFs are found not to be fundamentally different from ordinary flares \citep{Neidig1989, Hudson2006}. And their rarity in observation is thought to be primarily due to the limited sensitivity and resolution of telescopes \citep{Jess2008}. It is worth noting that the WL emission intensity does not exhibit a clear correlation with the GOES class of the flares, which is inconsistent with the expectations of the ``Big Flare Syndrome'' \citep{Kahler1982}. This may imply distinct types of WLFs, with significantly different  WL emission mechanisms \citep{Machado1986, Fang1995, Ding1999b}.

Clearly, we still do not know how common WLFs are on the Sun, and our picture of solar flares remains incomplete. Addressing this question requires both highly sensitive detection techniques and sufficient, stable, high-resolution observational data. At present, WLF detection is largely based on imaging observations. The traditional method is making difference maps between two times and then applying a threshold to identify WL enhancements. However, the WL emission at different locations is not strictly synchronized, this method cannot simultaneously detect all the WL kernels in a flare. It is therefore necessary to perform independent detection for each pixel in the target regions. Recently, we developed a new method to identify WLFs based on pixel-by-pixel light curve diagnostics \citep{Lishuanghong2026}. The method demonstrates exceptional detection capability. We apply it to an in-depth analysis of an X2.2 flare, as well as to a systematic detection of all flares (1 X-class, 2 M-class, and 20 C-class) from a typical active region (AR). We found a wealth of evolutionary details in WLFs and detected extremely weak WL enhancements. Details of the method are presented in the Methods section.

To reliably address this long-standing debate, a complete sample of flares is required \citep{Hudson2016}. Currently, the available WL continuum datasets for flares are derived from a variety of instruments, including space-based missions such as SDO/HMI, ASO-S/WST, and CHASE, as well as ground-based observatories like the Swedish 1-m Solar Telescope \citep[SST;][]{Scharmer2003},  the Optical and Near-infrared Solar Eruption Tracer\citep[ONSET;][]{Fang2013}, and the Accurate Infrared Magnetic field Measurements of the Sun \citep[AIMS;][]{Dengyy2025}. Ground-based observations are often affected by weather and seeing conditions, resulting in discontinuous and less stable data, and consequently capturing only a limited number of flares. Among the space-based datasets, SDO/HMI has been operating for over 15 years, covering the most of Solar Cycle 24 and a large part of Solar Cycle 25. Therefore, considering data stability, continuity, quality and temporal coverage, HMI observations offer significant advantages.

Since the HMI continuum is reconstructed from six wavelength sampling positions across the spectral line of Fe {\footnotesize I} 6173 \AA\ \citep{Couvidat2012}, it is thought that the spectral line undergoes significant perturbations during flares and the obtained continuum will be affected by spectral line core emission \citep{Mravcova2017,Svanda2018}. However, recent studies have shown that even in very small C-class WLFs or for weak WL emission in other flares, the radiation at all six HMI sampling positions exhibits clear enhancements, suggesting that the continuum enhancement is not solely due to line-core emission \citep{Song2020, Song2026, Jing2026}. Moreover, comparative studies have shown that the HMI continuum enhancement is weaker and its area is smaller than those measured at other wavelengths \citep[e.g.][]{SongDC2023, Liying2024a, Yangxu2025a, Yangxu2025b}. Notably, recent RADYN simulations have demonstrated that the HMI continuum enhancement can effectively reflect the heating of the lower solar atmosphere—a process that is fundamental to the generation of WLFs \citep{Hong2018, Granovsky2025}. Based on the above findings, the HMI continuum can be used for WLF detection. We therefore selected the HMI dataset spanning more than 15 years as the sample source for this study.

Up to now, only a few statistical studies have been devoted to the occurrence rate of WLFs. However, these works suffer from various limitations: they have either focused on large X- and M-class flares  \citep[e.g.][]{Neidig1983, Cai2024, Jing2024}, concentrated on specific types such as circular-ribbon or compact flares \citep{Song2018c, Song2026}, or targeted only extremely special ARs such as NOAA 11515, 13664/13697 \citep[e.g.][]{Song2018b, Liying2024b, Jing2025}. Consequently, these statistical results can only reflect WLF occurrence under particular ARs, specific magnetic configurations, or certain energy release conditions, and thus fail to provide a true picture of the overall occurrence of WLFs on the Sun. However, given the vast number of solar flares, ranging from A- to X-class, it is impractical to examine every single flare. We therefore plan to focus on flares from ARs with a low level of activity (flare number $<$ 10), observed during January 2011 -- May 2025. This sample selection strategy offers two advantages: 1) these  low-activity ARs represent the lower end of solar activity, allowing us to probe the lower limit for WLF production; 2) it substantially reduces the sample size and effectively lowers the analytical workload.

At last, we investigate a total of 1951 flares from 769 low-activity ARs during January 2011 – May 2025. The sample comprises 547 B-class, 1316 C-class, and 88 M-class flares. This study is the first systematic investigation of WLF occurrence based on over 15 years of HMI imaging observations, aiming to provide a reliable evaluation of how common WLFs are on the Sun. More than 170 years after Carrington’s first observation of WLF, and on the 200th anniversary of his birth (1826--2026), we expect this work to bring new insights into the nature of WLFs.

\section{Data and observations}\label{sec2}

\begin{table*}[htbp]
\centering
\caption{Sample of low-activity active regions (flare number $<$ 10)  from January 2011 to May 2025}
\label{tab1}
{\fontsize{6.8}{8.8}\selectfont
\begin{tabular}{c|l|ccccccccc}
\hline
\hline
\multicolumn{1}{c}{Year} & \multicolumn{1}{c}{NOAA Number} & \multicolumn{1}{c}{Total}\\
\hline
	\multirow{5}{*}{2011} & 11142, 11152, 11156, 11157, 11159, 11173, 11178, 11181, 11186, 11196, 11200, 11201, 11202, 11203, 11204 & \multirow{5}{*}{72}  \\
	       & 11208, 11210, 11214, 11217, 11218, 11223, 11224, 11227, 11228, 11229, 11232, 11235, 11237, 11239, 11240  & \\
	       & 11241, 11242, 11245, 11249, 11255, 11257, 11262, 11267, 11269, 11274, 11281, 11289, 11301, 11309, 11315 & \\
	       & 11316, 11317, 11318, 11325, 11330, 11333, 11334, 11338, 11341, 11342, 11347, 11353, 11357, 11358, 11361 & \\
	       & 11362, 11363, 11364, 11366, 11368, 11374, 11375, 11377, 11381, 11382, 11384, 11386 & \\
\hline
       \multirow{5}{*}{2012} & 11391, 11392, 11395, 11396, 11403, 11413, 11417, 11418, 11421, 11423, 11427, 11428, 11430, 11433, 11438 & \multirow{5}{*}{70}  \\
	       & 11440, 11442, 11444, 11446, 11452, 11455, 11460, 11462, 11466, 11467, 11470, 11473, 11474, 11479, 11482  & \\
	       & 11483, 11484, 11488, 11490, 11491, 11492, 11496, 11498, 11506, 11512, 11514, 11522, 11524, 11525, 11526 & \\
	       & 11528, 11536, 11540, 11544, 11553, 11555, 11562, 11565, 11567, 11568, 11573, 11577, 11578, 11579, 11582 & \\
	       & 11584, 11590, 11591, 11597, 11608, 11611, 11615, 11628, 11637, 11638 & \\
\hline
	\multirow{6}{*}{2013} & 11643, 11645, 11657, 11658, 11661, 11663, 11665, 11669, 11673, 11676, 11680, 11681, 11682, 11689, 11690 & \multirow{6}{*}{79}  \\
	       & 11691, 11695, 11696, 11700, 11704, 11708, 11710, 11711, 11715, 11721, 11723, 11724, 11732, 11733, 11736  & \\
	       & 11738, 11742, 11743, 11750, 11752, 11753, 11754, 11755, 11757, 11761, 11768, 11769, 11773, 11777, 11780 & \\
	       & 11781, 11784, 11788, 11801, 11803, 11805, 11810, 11813, 11819, 11825, 11827, 11828, 11838, 11845, 11846 & \\
	       & 11856, 11857, 11868, 11870, 11873, 11885, 11887, 11889, 11891, 11907, 11908, 11909, 11916, 11920, 11921 & \\
	       & 11925, 11926, 11930, 11931 & \\
\hline	       
	\multirow{6}{*}{2014} & 11937, 11946, 11950, 11953, 11955, 11956, 11957, 11960, 11972, 11973, 11975, 11977, 11980, 11984, 11986 & \multirow{6}{*}{81}  \\   
	       & 11987, 11989, 11990, 11993, 12005, 12008, 12013, 12021, 12022, 12024, 12027, 12029, 12031, 12032, 12034 & \\    
	       	& 12037, 12044, 12045, 12049, 12050, 12053, 12055, 12058, 12059, 12063, 12066, 12069, 12071, 12072, 12073 & \\  
		& 12077, 12082, 12092, 12100, 12107, 12121, 12122, 12123, 12125, 12126, 12127, 12134, 12135, 12137, 12140 & \\   
		& 12141, 12142, 12147, 12148, 12150, 12151, 12164, 12175, 12176, 12177, 12178, 12181, 12185, 12214, 12219 & \\  
		& 12221, 12226, 12234, 12235, 12245, 12251 & \\  
\hline	
      \multirow{4}{*}{2015} & 12258, 12260, 12262, 12266, 12271, 12273, 12276, 12281, 12284, 12287, 12293, 12294, 12310, 12312, 12318 & \multirow{4}{*}{51}  \\
              & 12323, 12327, 12331, 12338, 12342, 12345, 12349, 12351, 12352, 12353, 12356, 12362, 12364, 12365, 12366 & \\
              & 12376, 12385, 12386, 12387, 12393, 12394, 12400, 12405, 12409, 12410, 12414, 12421, 12432, 12440, 12448 & \\
              & 12449, 12459, 12462, 12464, 12466, 12469 & \\
\hline
      \multirow{3}{*}{2016} & 12476, 12479, 12482, 12483, 12485, 12487, 12490, 12493, 12496, 12500, 12501, 12503, 12505, 12511, 12512 & \multirow{3}{*}{45}  \\
             & 12515, 12519, 12521, 12523, 12528, 12531, 12532, 12540, 12541, 12548, 12549, 12550, 12553, 12554, 12562 & \\
             & 12576, 12578, 12579, 12588, 12589, 12592, 12593, 12599, 12602, 12603, 12610, 12612, 12613, 12614, 12616 & \\
\hline
      \multirow{2}{*}{2017} & 12626, 12636, 12640, 12647, 12648, 12653, 12654, 12655, 12656, 12657, 12658, 12660, 12680, 12681, 12687 & \multirow{2}{*}{17}  \\
            & 12689, 12690 & \\
\hline
      \multirow{1}{*}{2018} & 12696, 12697, 12710, 12719, 12724, 12728, 12731 & \multirow{1}{*}{7}  \\
 \hline     
      \multirow{1}{*}{2019} & 12734, 12735, 12741 & \multirow{1}{*}{3}  \\
 \hline         
      \multirow{1}{*}{2020} & 12757, 12759, 12765, 12772, 12773, 12780, 12794 & \multirow{1}{*}{7}  \\
 \hline  
      \multirow{2}{*}{2021} & 12800, 12808, 12810, 12812, 12819, 12821, 12823, 12830, 12836, 12837, 12839, 12841, 12842, 12857, 12858 & \multirow{2}{*}{29}  \\
            & 12861, 12864, 12872, 12873, 12875, 12883, 12889, 12890, 12897, 12904, 12909, 12915, 12917, 12920& \\
 \hline  
      \multirow{6}{*}{2022} & 12925, 12926, 12927, 12930, 12935, 12943, 12946, 12948, 12950, 12955, 12956, 12958, 12959, 12961, 12968 & \multirow{6}{*}{80}  \\
            & 12970, 12971, 12972, 12976, 12983, 12985, 12987, 12995, 12996, 12997, 13001, 13015, 13016, 13019, 13021 & \\  
            & 13024, 13026, 13029, 13032, 13035, 13037, 13046, 13047, 13050, 13051, 13052, 13060, 13062, 13063, 13065 & \\
            & 13067, 13073, 13074, 13076, 13077, 13082, 13083, 13084, 13092, 13093, 13096, 13097, 13099, 13103, 13109 & \\
            & 13111, 13113, 13116, 13118, 13131, 13133, 13136, 13137, 13139, 13145, 13148, 13152, 13155, 13156, 13158 & \\
            & 13161, 13166, 13171, 13172, 13174  & \\      
  \hline  
      \multirow{6}{*}{2023} & 13183, 13185, 13191, 13194, 13196, 13199, 13206, 13211, 13214, 13220, 13230, 13235, 13236, 13241, 13250 & \multirow{6}{*}{78}  \\   
            & 13258, 13259, 13262, 13263, 13265, 13273, 13283, 13285, 13289, 13294, 13299, 13306, 13309, 13314, 13316 & \\        
            & 13319, 13324, 13330, 13331, 13334, 13336, 13338, 13339, 13348, 13351, 13367, 13370, 13374, 13377, 13379 & \\        
            & 13382, 13390, 13392, 13397, 13403, 13409, 13415, 13422, 13431, 13437, 13447, 13449, 13453, 13454, 13457 & \\        
            & 13463, 13465, 13467, 13468, 13471, 13479, 13481, 13486, 13494, 13499, 13502, 13503, 13507, 13509, 13521 & \\ 
            & 13524, 13533, 13534 & \\       
  \hline             
      \multirow{7}{*}{2024} & 13542, 13545, 13546, 13547, 13548, 13560, 13565, 13567, 13574, 13582, 13584, 13586, 13587, 13588, 13591 & \multirow{7}{*}{104}  \\   
            & 13594, 13595, 13596, 13598, 13600, 13602, 13603, 13604, 13607, 13613, 13616, 13617, 13628, 13630, 13632 & \\  
            & 13633, 13635, 13636, 13637, 13641, 13643, 13646, 13648, 13672, 13673, 13674, 13686, 13695, 13698, 13701 & \\
            & 13703, 13711, 13714, 13716, 13718, 13719, 13720, 13727, 13728, 13730, 13731, 13733, 13736, 13740, 13743 & \\
            & 13745, 13750, 13753, 13754, 13757, 13761, 13764, 13770, 13775, 13779, 13782, 13785, 13793, 13794, 13799 & \\
            & 13805, 13808, 13828, 13831, 13835, 13843, 13849, 13850, 13859, 13860, 13863, 13866, 13872, 13873, 13875 & \\
            & 13877, 13881, 13884, 13890, 13891, 13892, 13893, 13898, 13908, 13909, 13910, 13911, 13925, 13940 & \\
  \hline             
      \multirow{4}{*}{2025} & 13943, 13945, 13950, 13951, 13963, 13965, 13967, 13969, 13971, 13993, 13994, 13997, 14000, 14009, 14010 & \multirow{4}{*}{46}  \\  
           & 14011, 14016, 14018, 14020, 14022, 14025, 14027, 14031, 14033, 14034, 14035, 14041, 14044, 14045, 14046 & \\
           & 14049, 14052, 14057, 14062, 14063, 14065, 14067, 14073, 14075, 14081, 14085, 14089, 14094, 14097, 14101 & \\
           & 14104 & \\
 \hline 
	\multicolumn{1}{c}{Total} & \multicolumn{1}{c}{ }  & \multicolumn{1}{c}{769}  \\													
\hline
\end{tabular}
}
\end{table*}

This study primarily uses data from the HMI and the Atmospheric Imaging Assembly \citep[AIA;][]{Lemen2011} onboard SDO. HMI observes the Sun in the Fe {\footnotesize I} 6173~\AA\ absorption line and provides data products including continuum intensity maps, Dopplergrams, line-of-sight (LOS) magnetograms, and vector magnetograms. HMI has a pixel scale of $\sim$$0.5^{\prime\prime}$ and a spatial resolution of about $1^{\prime\prime}$. Different data products have different cadences. In this study, we mainly use the 45~s HMI continuum intensity maps to analyze the WL emission during the flare, and combine them with the 720~s vector magnetic field data to examine the relationship between the WL enhancement and the local photospheric magnetic field strength.

AIA observes the Sun in multiple extreme-ultraviolet (EUV) and ultraviolet (UV) passbands, which can be used to study solar atmospheric structures at different layers. The EUV passbands include 94~\AA, 131~\AA, 171~\AA, 193~\AA, 211~\AA, 304~\AA, and 335~\AA, covering a temperature range of approximately $10^5$--$10^7$ K. The UV passbands are 1600~\AA\ and 1700~\AA. AIA has a pixel scale of $\sim$0.6$^{\prime\prime}$ and a spatial resolution of about 1.2$^{\prime\prime}$. The temporal resolutions for EUV and UV observations are 12~s and 24~s, respectively. In this study, we use the AIA 1600~\AA\ images to identify the flare regions. 

We also use the 1--8~\AA\ soft X-ray (SXR) flux data provided by the Geostationary Operational Environmental Satellites \citep[GOES;][]{Hanser1996} to determine the start, peak, and end times of the flare, and to characterize its overall temporal evolution. 

Table \ref{tab1} shows the 769 low-activity ARs from January 2011 to May 2025. These low-activity ARs are defined by a flare count of fewer than 10 during their disk passage. The flare numbers in each AR are retrieved from the GOES X-ray flare event list\footnote{\url{https://hesperia.gsfc.nasa.gov/goes/goes_event_listings/}} and SpaceWeatherLive \footnote{\url{https://www.spaceweatherlive.com}}. After further excluding flares located near the solar limb, we ultimately obtained 1951 flares from these ARs. The class distribution of the flare sample is summarized in Table \ref{tab2}, comprising 1316 C-class, 547 B-class, and 88 M-class flares. Overall, the sample is dominated by C- and B-class flares, with a relatively small number of M-class flares and no X-class flares.

\begin{table*}[htbp]
\centering
\caption{Annual distribution of the flare sample from January 2011 to May 2025}
\label{tab2}
{\fontsize{8.0}{11.2}\selectfont
\begin{tabular}{ccccccccccc}
\hline
\hline
Year & B-class flares & C-class flares & M-class flares & Total  \\
\hline
	2011 & 68 & 135 & 8 & 206  \\
	2012 & 41 & 152 & 6  & 199  \\
	2013 & 56 & 135 & 2  & 193   \\
	2014 & 33 & 182 & 8  & 223   \\
	2015 & 46 & 73  & 2  & 121   \\
	2016 & 94 & 33  & 0  & 127   \\
	2017 & 39 & 2   & 0  & 41   \\
	2018 & 9  & 3   & 0  & 12    \\
	2019 & 13 & 2   & 0  & 15    \\
	2020 & 13 & 1   & 0  & 14   \\
	2021 & 46 & 22  & 0  & 68   \\
	2022 & 80 & 126 & 7  & 213  \\
	2023 & 2  & 167 & 20 & 189 \\
	2024 & 4  & 191 & 31 & 226  \\
	2025 & 3  & 92  & 9  & 104  \\
\hline
	Total & 547 & 1316 & 88 & 1951 \\
\hline
       	WLF counts & 202 & 735 & 75 & 1012 \\
	WLF occurrence rate & 36.93\% & 55.85\% & 85.23\% & 51.87\% \\
												
\hline
\end{tabular}
}
\end{table*}

\section{Methods}\label{sec3}

In this work, we detect WLFs using the method proposed by \citet{Lishuanghong2026}. Physically, WLFs arise from efficient heating of the lower solar atmosphere, which generates enhanced continuum emission. Such brightening signals typically exhibit pronounced pulsed features. The core idea of this method is to compare the amplitude of WL continuum emission pulses during flares with those measured before and after the flare at each pixel, so as to identify whether WL brightening emerges at that location.

The specific procedure is as follows. First, we identify the flare regions using the AIA 1600 \AA\ observations. We then extract the light curve at each pixel in the flare regions from the HMI continuum data. The time interval of the light curve typically covers 10 minutes before the flare onset to 10 minutes after the flare end. Next, we identify a series of local minima along the light curve and construct the background by connecting these minima. Finally, after subtracting the background from the original light curve, we obtain a series of WL emission pulses. If the pulse intensity during the flare is significantly higher than that before and after the flare, the corresponding pixel is marked as a WL pixel. The threshold is set as $(I_{fm}-\frac{1}{N}\sum_{i=1}^{N} I_{0i})\ge6\sigma$. Here $I_{fm}$ is the maximum pulse intensity during the flare, $I_{0i}$ is the pulse intensity before and after the flare, $N$ is the total number of these pre- and post-flare pulses, and $\sigma$ is the standard deviation of these pulses. It should be noted that, to exclude interference from random heating events, hot pixels, and cosmic rays, the method removes isolated WL points ($\le$ 2 pixels) to reduce the risk of false detections. 

This method is physically clear and operationally straightforward, requiring no complex algorithms. It can be widely applied to flares of various classes and types, and demonstrates excellent detection capability even for very weak WLFs \citep{Lishuanghong2026}, making it well suited for large-sample WLF detection and statistical studies. Moreover, by independently examining the light curve of each pixel, it can reveal rich evolutionary details of WLFs, which is of great value for understanding the flare process and the heating mechanisms of the lower atmosphere. Crucially, its sensitivity to weak WL emission, combined with our flare sample dominated by B- and C-class flares from low-activity ARs, will significantly advance our understanding of the prevalence of WLFs on the Sun.

\section{Results}\label{sec4}

\begin{figure}[h]
\centering
\includegraphics[trim=0.0cm 1.1cm 0.0cm 0.5cm, width=1.02\textwidth]{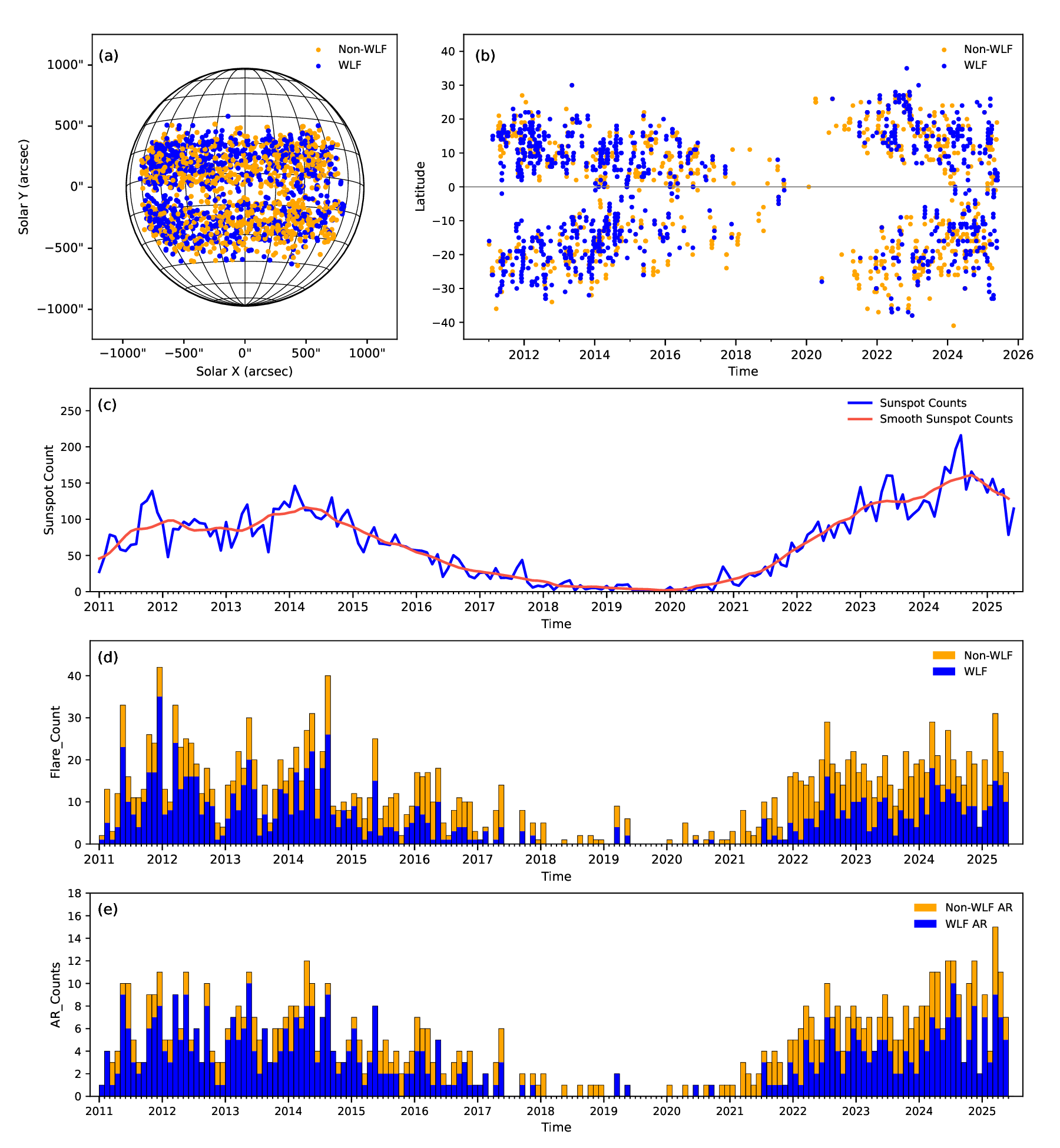}
\caption{
Spatial and temporal distributions of the flare sample from low-activity active regions (ARs; flare number $<$ 10) during January 2011 - May 2025. (a) Two-dimensional spatial distribution of the WLFs (blue) and non-WLFs (yellow) on the solar disk in helioprojective coordinates. (b) Temporal evolution of flare latitudes, presented as a butterfly diagram. (c) Monthly mean sunspot number (blue) and its 13-month smoothed curve (red). (d) Monthly counts of the WLFs (blue) and non-WLFs (yellow). (e) Monthly counts of the low-activity ARs with (blue) and without (yellow) WLFs..
}\label{fig1}
\end{figure}

\begin{figure}[h]
\centering
\includegraphics[width=0.98\textwidth]{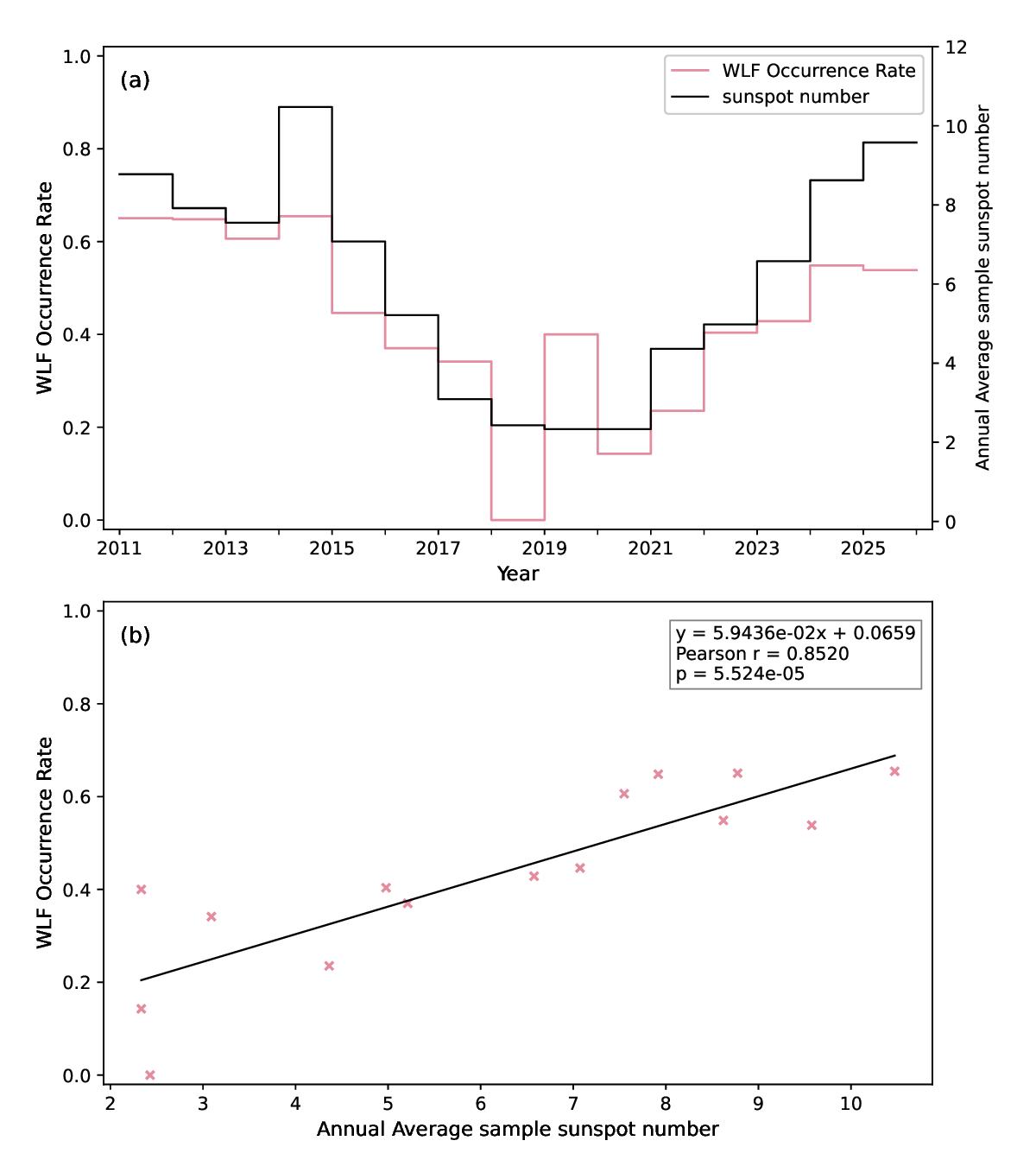}
\caption{
Relationship between the WLFs occurrence rate and the yearly mean sunspot number of the sample ARs. (a) Yearly variations of the WLFs occurrence rate (pink step line) and the yearly mean sunspot number of the sample ARs (black step line) during 2011--2025. (b) Scatter plot and linear fit between the WLFs occurrence rate and the yearly mean sunspot number of the sample ARs. The pink crosses indicate the yearly statistical results, and the black solid line represents the linear fit.
}\label{fig2}
\end{figure}

\begin{figure}[h]
\centering
\includegraphics[width=0.98\textwidth]{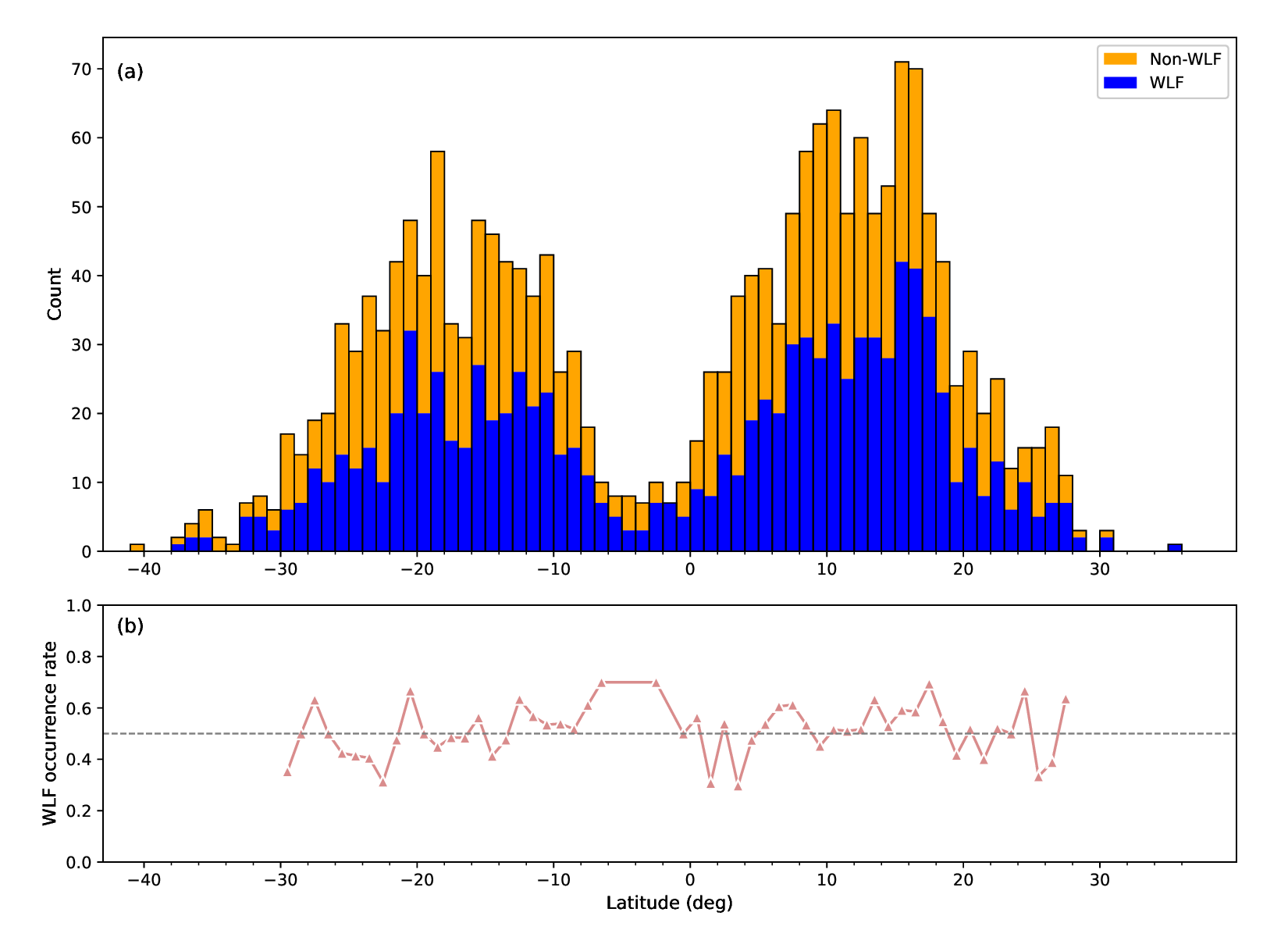}
\caption{
Latitudinal distributions of WLFs and non-WLFs. (a) Numbers of WLFs and non-WLFs in different latitude bins. (b) The corresponding WLF occurrence rates. The gray dashed line indicates an occurrence rate of 50\%
}\label{fig3}
\end{figure}

Figure \ref{fig1} shows the spatial distribution of these WLFs and non-WLFs from low-activity ARs (flare number $<$ 10), as well as their temporal evolution across Solar Cycles 24 and 25 between January 2011 and May 2025. As seen in panel (a), both WLFs and non-WLFs appear to be approximately randomly distributed across the solar disk, with no obvious longitudinal or latitudinal preference. It should be noted that, to avoid projection effects, we excluded events near the solar limb and considered only those with longitudes within $\pm 60^\circ$. Panel (b) further shows the butterfly diagram distribution of these events, in which WLFs clearly follow a typical butterfly pattern, indicating a close association with the solar cycle. Panel (c) displays the sunspot number distribution from January 2011 to May 2025, covering most of Solar Cycle 24 and a large part of Solar Cycle 25. Panels (d) and (e) show the temporal variations of flare numbers and low-activity AR numbers over the solar cycle, respectively. Both are consistent with the solar cycle variation, further demonstrating that our statistical sample reflects the overall characteristics of the solar cycle. We find that, in both solar cycles, approximately half of these flares are WLFs. Statistical results show that the fraction of WLFs is 56.90\% in Solar Cycle 24 and 44.84\% in Solar Cycle 25, with an overall occurrence rate of 51.87\%. For the low-activity ARs, more than half contain WLFs: 72.24\% in Solar Cycle 24 and 58.43\% in Solar Cycle 25, with an overall fraction of 66.10\%. This indicates that even in relatively low-activity ARs, WLFs occur quite commonly, supporting that WLFs are ubiquitous on the Sun.

Figure \ref{fig2} shows the relationships between the WLF occurrence rate and the yearly mean sunspot number of the sample ARs. Here, the yearly mean sunspot number of the sample ARs, $\bar{S}_y$, is defined as the annual average of the daily total sunspot number of all sample ARs over all valid sample days in year $y$, expressed as
\begin{equation}
	\bar{S}_y = \frac{1}{D_y}\sum_{d=1}^{D_y}\sum_{i=1}^{N_{y,d}} n_{y,d,i}
	\label{eq:sunspot_avg}
\end{equation}
where $y$ denotes the year, $D_y$ is the number of valid sample days in year $y$, $N_{y,d}$ is the number of sample ARs on day $d$ of year $y$, and $n_{y,d,i}$ is the sunspot number of the $i$-th sample AR on that day. The AR sunspot numbers were obtained from the NOAA/SWPC Solar Region Summary daily report archive\footnote{\url{https://www.ngdc.noaa.gov/stp/space-weather/swpc-products/daily_reports/solar_region_summaries/}}. From panel (a), we see the two quantities show good consistency in their overall temporal variations: when the yearly mean sunspot number of the sample ARs is relatively high, the WLF occurrence rate is usually also high; conversely, in years with lower yearly mean sunspot numbers, the occurrence rate also tends to be lower. In particular, around the solar minimum, the yearly mean sunspot number decreases to a low level, and the WLF occurrence rate is also generally low.

Panel (b) shows the scatter distribution and linear fit between the WLF occurrence rate and the yearly mean sunspot number of the sample ARs. The statistical analysis shows a strong positive correlation between the two quantities, with a Pearson correlation coefficient of $r\approx0.85$, indicating that the WLF occurrence rate generally increases with the yearly mean sunspot number of the sample ARs. The number of sunspots in an AR can, to some extent, characterize its magnetic complexity. In general, sunspot groups with more spots tend to have more complex magnetic structures and higher activity levels. In addition, the sunspot number is commonly used as an indicator of the photospheric magnetic activity level \citep{YanXL2012}. This indicates that WLF occurrence is closely associated with the level of solar magnetic activity.

Figure \ref{fig3} presents the number distributions of WLFs and non-WLFs in different latitude bins, together with the corresponding WLFs occurrence rates. To reduce the influence of small-number statistical fluctuations, the WLF occurrence rate was calculated only for latitude bins containing more than 10 flares. Panel (a) shows that the northern hemisphere hosts a larger number of flares and, correspondingly, a larger number of WLFs. However, panel (b) shows that the WLF occurrence rates in different latitude bins are mostly distributed around 50\%, with no clear monotonic variation with increasing or decreasing latitude. The occurrence rates in corresponding latitude ranges of the northern and southern hemispheres are also broadly comparable, showing no obvious north--south asymmetry. This result indicates that the WLF occurrence rate shows no clear latitudinal dependence.

\begin{figure}[h]
\centering
\includegraphics[trim=0.0cm 0.6cm 0.0cm 0.5cm, width=0.98\textwidth]{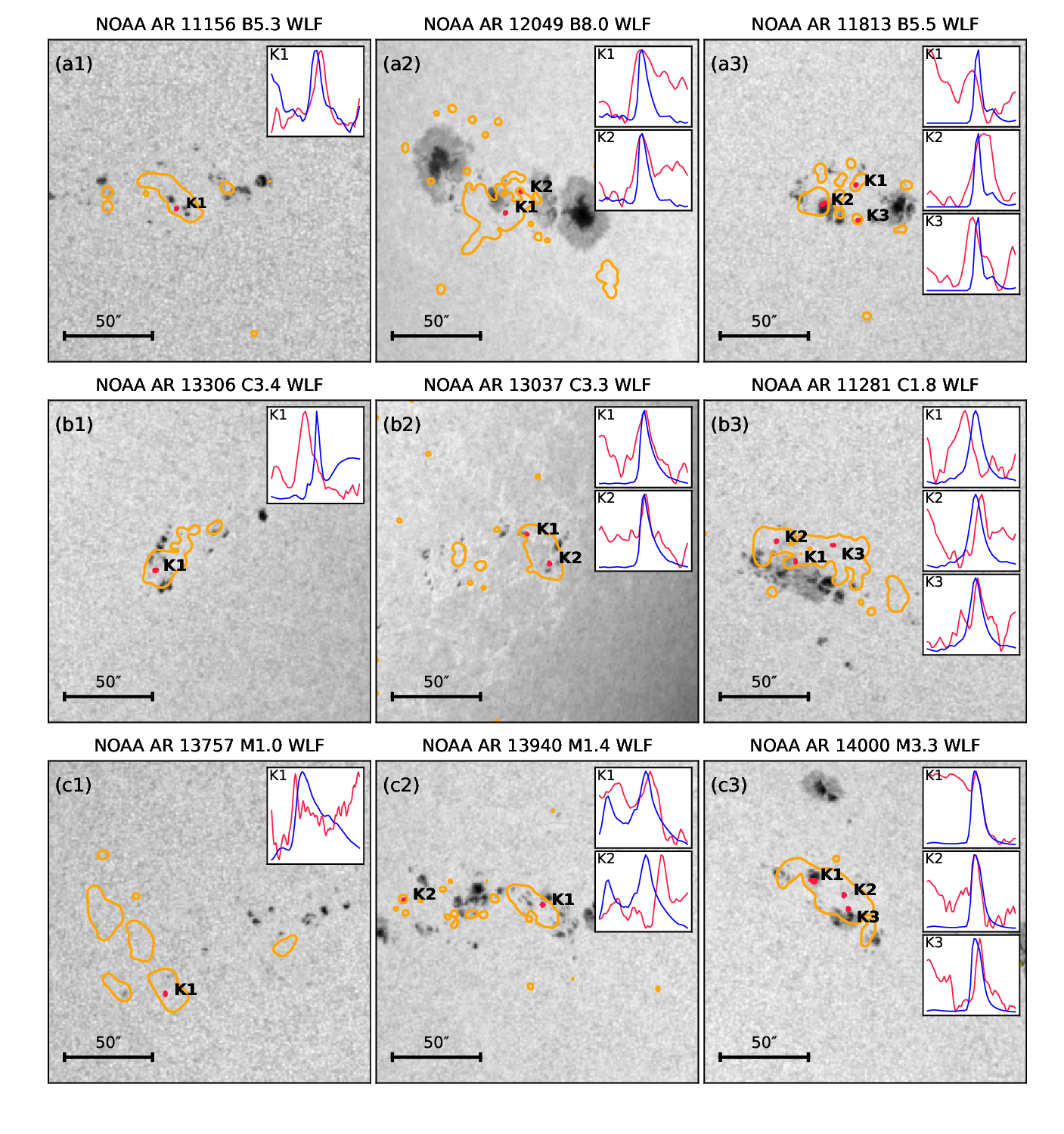}
\caption{
WLF sample with different GOES classes. (a1)–(a3) show the HMI continuum intensity maps of three B-class WLFs. The orange contours indicate the flare regions obtained from the AIA 1600 \AA\ observations, and the red areas denote the identified WL kernels. The number of WL kernels varies for different flares. The normalized HMI continuum light curves of each WL kernel (red lines) and the normalized GOES SXR flux for each flare (blue lines) are superimposed in each panel. Similarly, (b1)–(b3) and (c1)–(c3) show three C-class and three M-class WLFs, respectively.
}\label{fig4}
\end{figure}

\begin{figure}[h]
\centering
\includegraphics[trim=0.0cm 1.1cm 0.0cm 0.5cm, width=1.02\textwidth]{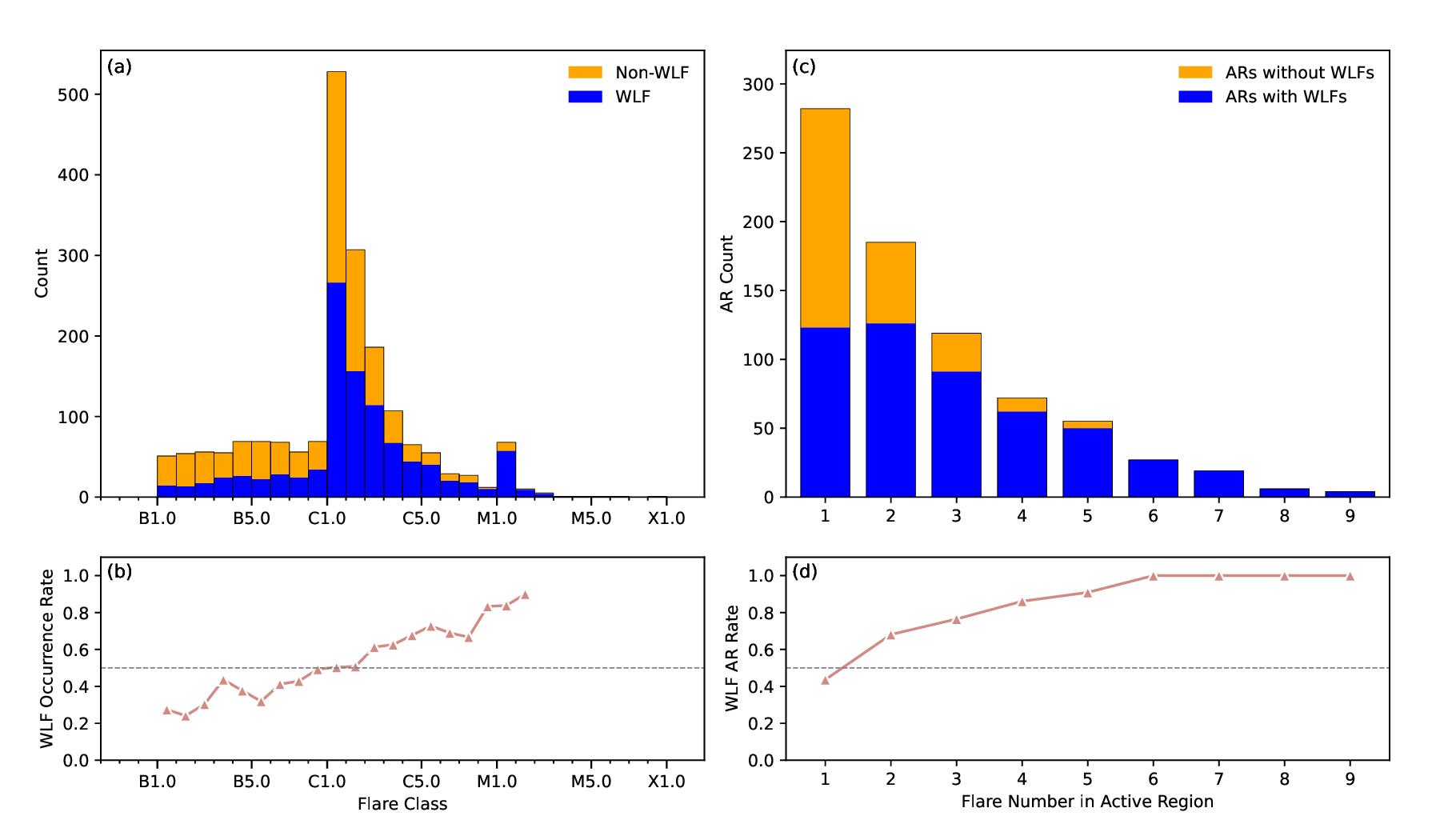}
\caption{
Number distributions and corresponding occurrence rates of WLFs and non-WLFs by flare GOES class, and of ARs with and without WLFs grouped by flare number. (a) Number distributions of WLFs (blue) and non-WLFs (orange) in different GOES-class bins. (b) WLF occurrence rates in different GOES-class bins. (c) Number distributions of ARs with (blue) and without (orange) WLFs, grouped by their flare number. (d) Fractions of ARs with WLFs in different flare-number groups. The gray dashed lines in panels (b) and (d) indicate an occurrence rate of 50\%.
}\label{fig5}
\end{figure}

\begin{figure}[h]
\centering
\includegraphics[trim=0.0cm 0.5cm 0.0cm 0.5cm, width=0.95\textwidth]{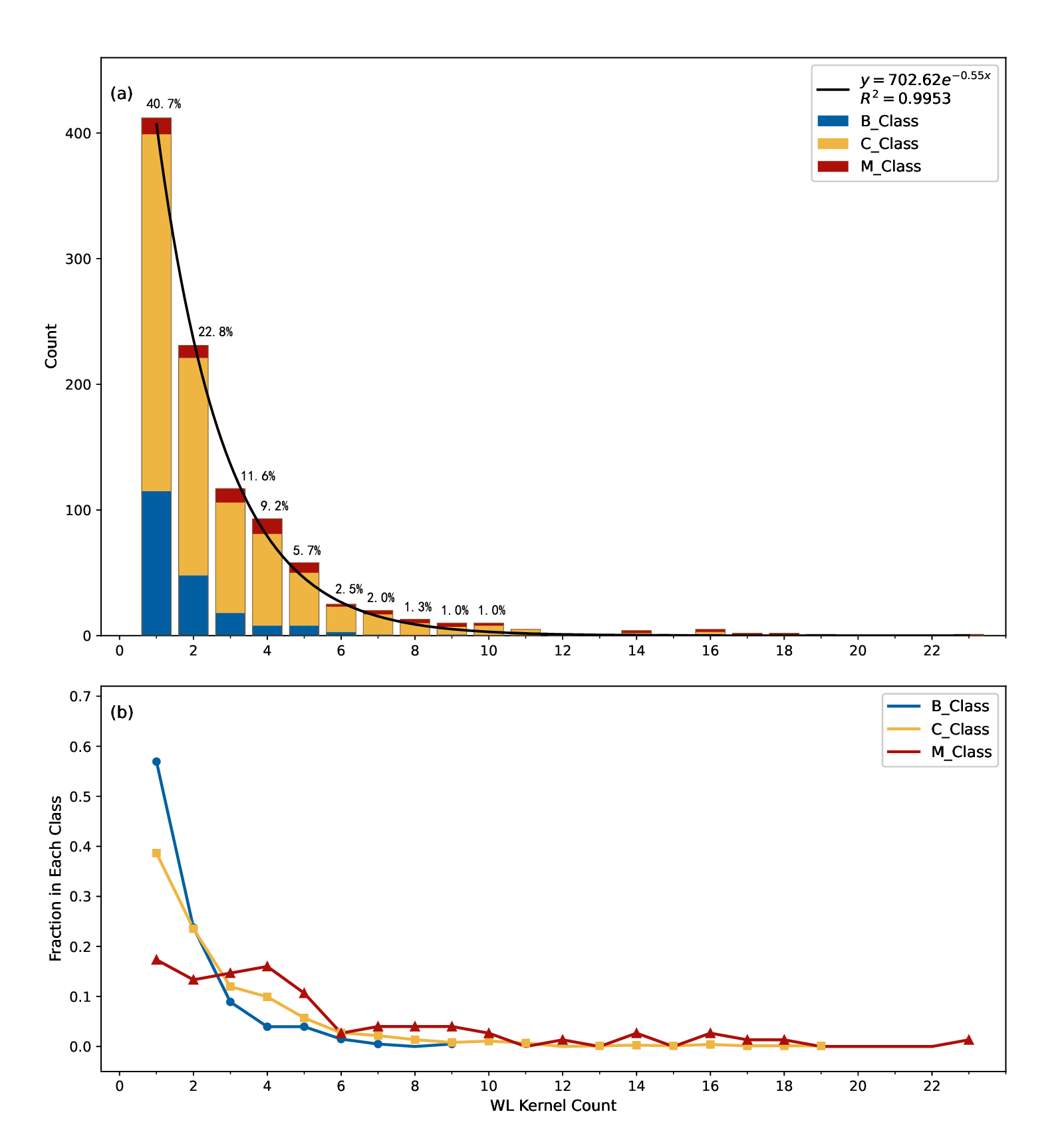}
\caption{
Distribution of WLFs with different numbers of WL kernels. (a) Histogram of the WLFs grouped by the number of WL kernels. The corresponding fractions are also given. Blue, orange, and red represent B-, C-, and M-class WLFs, respectively. The black solid line denotes the exponential fit. (b) Fractions of WLFs with different numbers of WL kernels for different flare classes, normalized by the total number of WLFs in each class.
}\label{fig6}
\end{figure}

\begin{figure}[h]
\centering
\includegraphics[trim=0.0cm 0.5cm 0.0cm 0.5cm, width=0.95\textwidth]{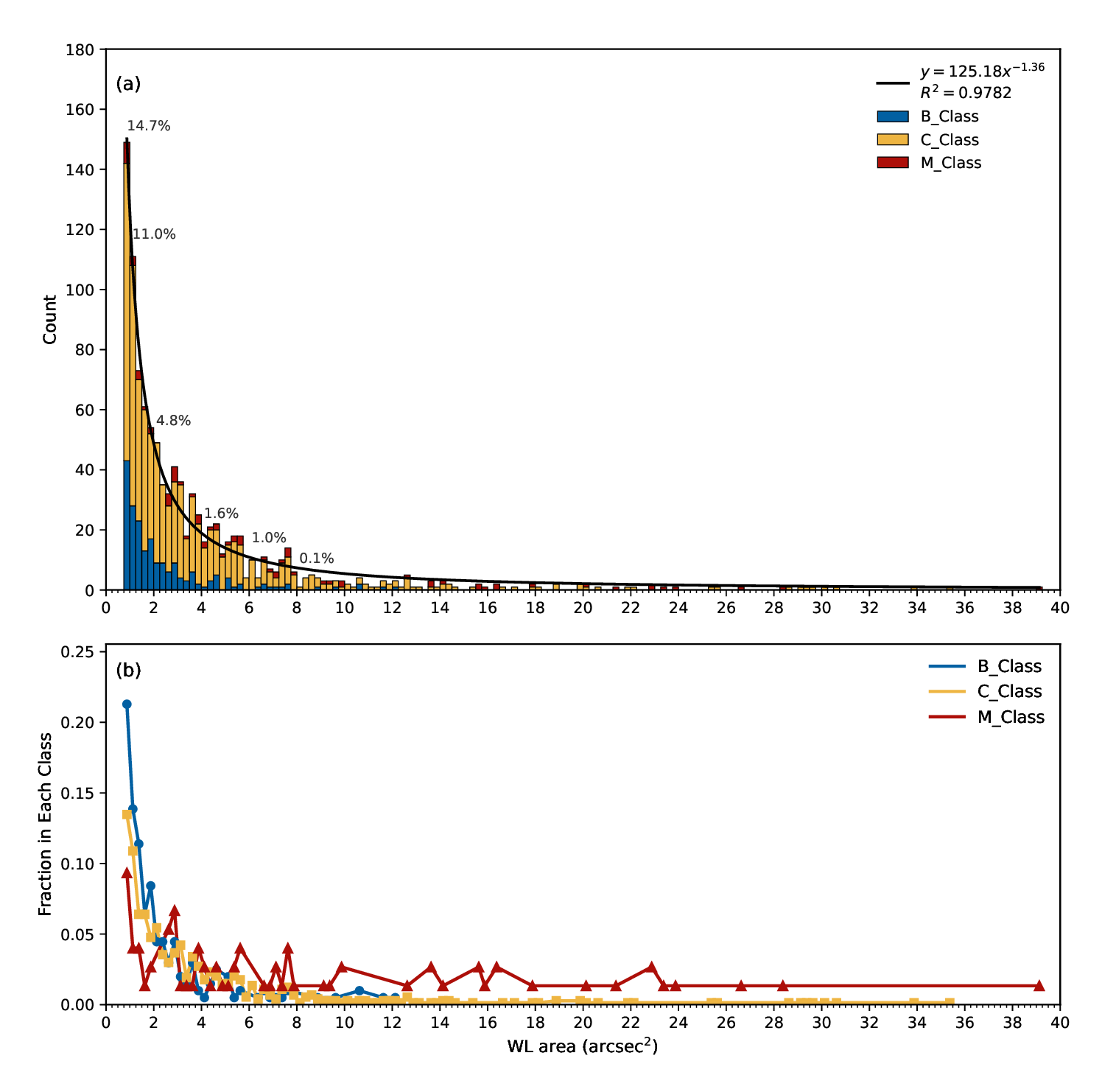}
\caption{
Same as Figure \ref{fig6}, but for distribution of WLFs with different WL areas. The black solid line in panel (a) denotes a power-law fit.
}\label{fig7}
\end{figure}

\begin{figure}[h]
\centering
\includegraphics[width=0.95\textwidth]{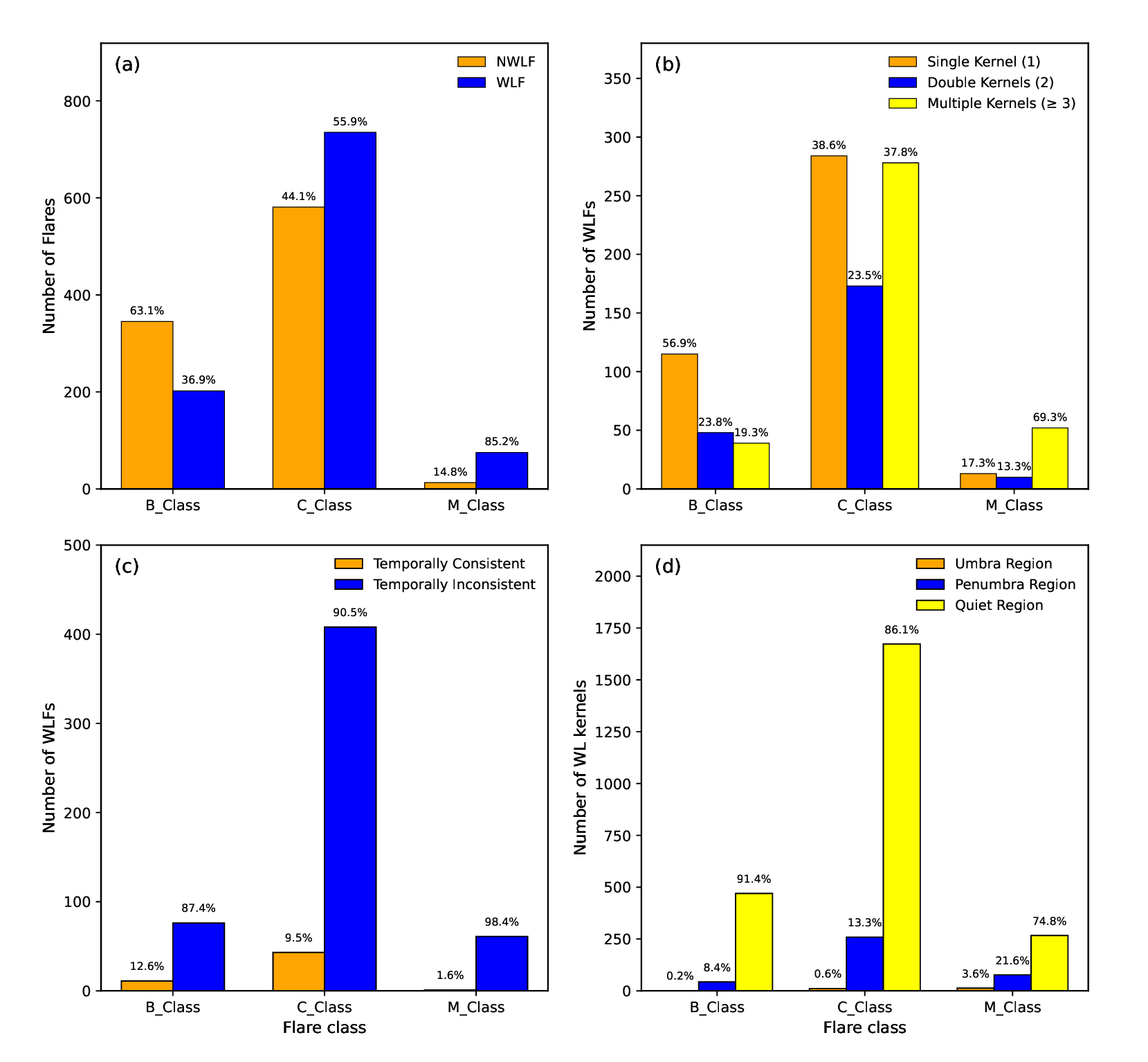}
\caption{
Statistics of WLFs and non-WLFs, and of the properties of WL kernels among B-, C- and M-class flares. (a) Percentages of WLFs and non-WLFs. (b) Percentages of WLFs with single, double, or multiple WL kernels. (c) Fractions of WLFs with temporally consistent or inconsistent WL kernels. (d) Distributions of WL kernel positions.
}\label{fig8}
\end{figure}

\begin{figure}[h]
\centering
\includegraphics[width=0.9\textwidth]{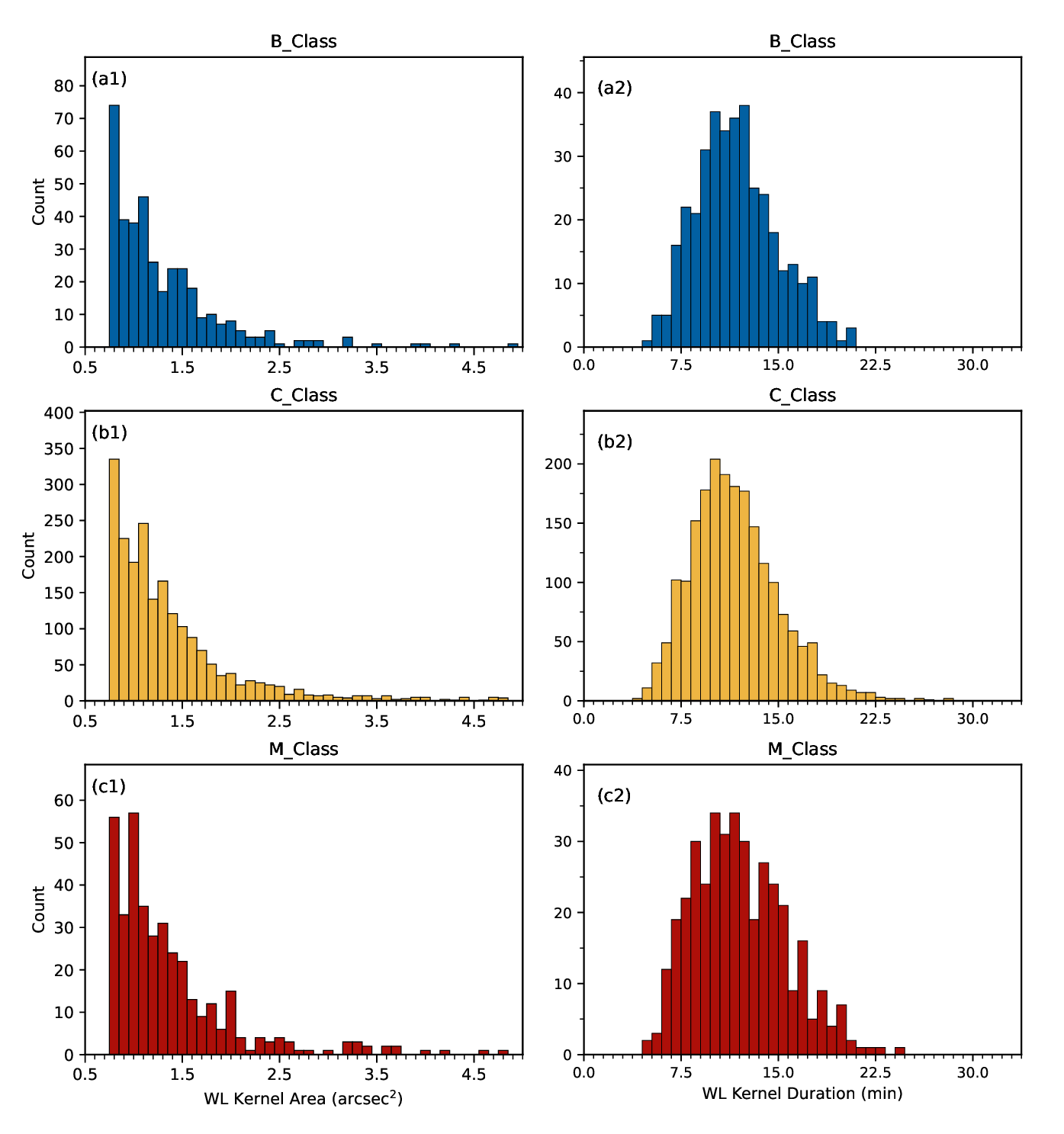}
\caption{
Distributions of the areas (a1, b1, c1) and durations (a2, b2, c2) of WL kernels for B-, C-, and M-class WLFs, respectively.
}\label{fig9}
\end{figure}

\begin{figure}[h]
\centering
\includegraphics[width=0.9\textwidth]{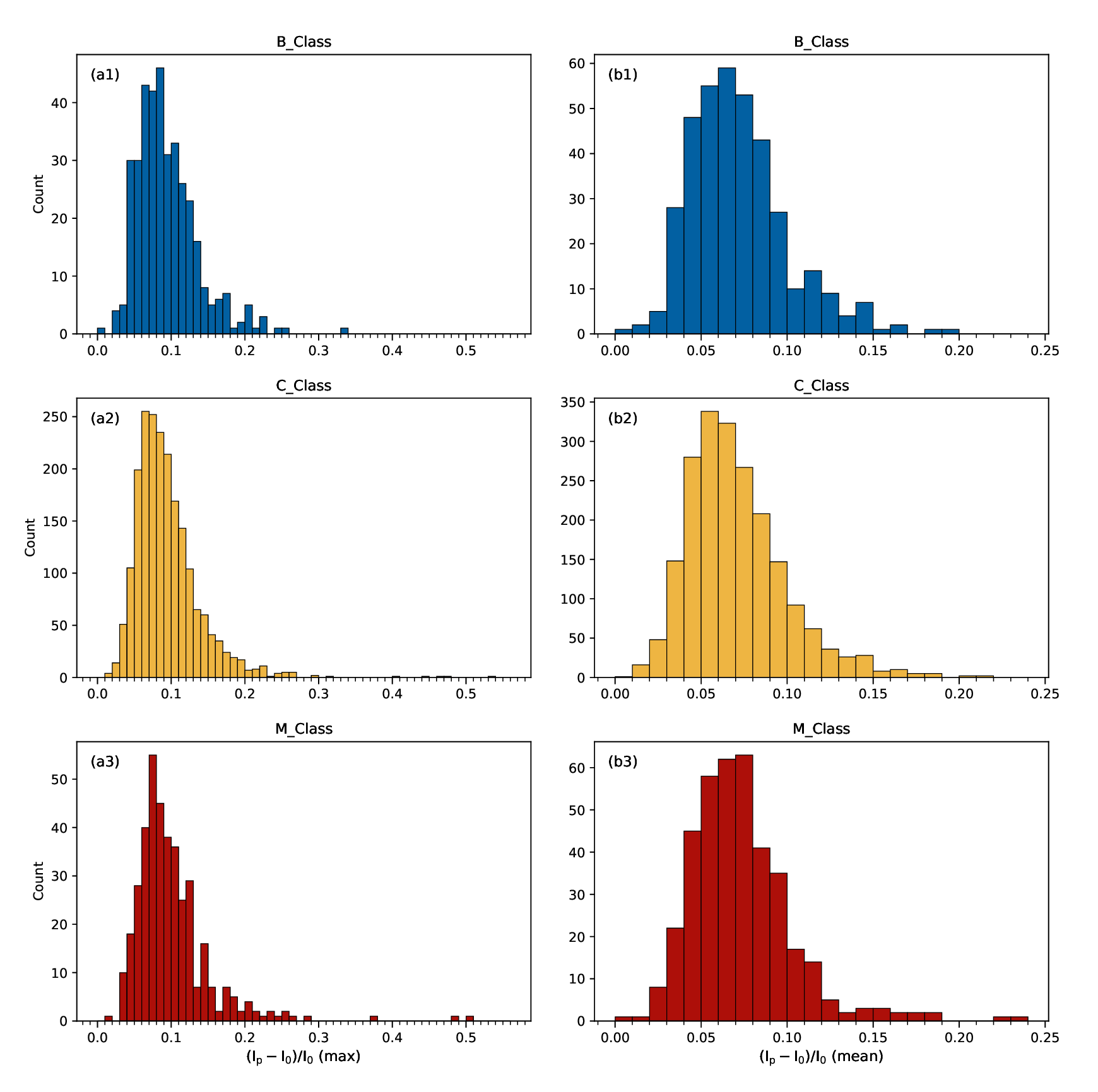}
\caption{
Histograms of the relative WL enhancements ($(I_p-I_0)/I_0$) of WL kernels in B-, C-, and M-class flares.  Here $I_p$ and $I_0$ denote the HMI continuum intensities at the WL emission peak and before the flare in a WL kernel. Panels (a1)–(a3) and (b1)–(b3) show the maximum and average relative WL enhancements in WL kernels for different GOES classes.
}\label{fig10}
\end{figure}

\begin{figure}[h]
\centering
\includegraphics[width=0.9\textwidth]{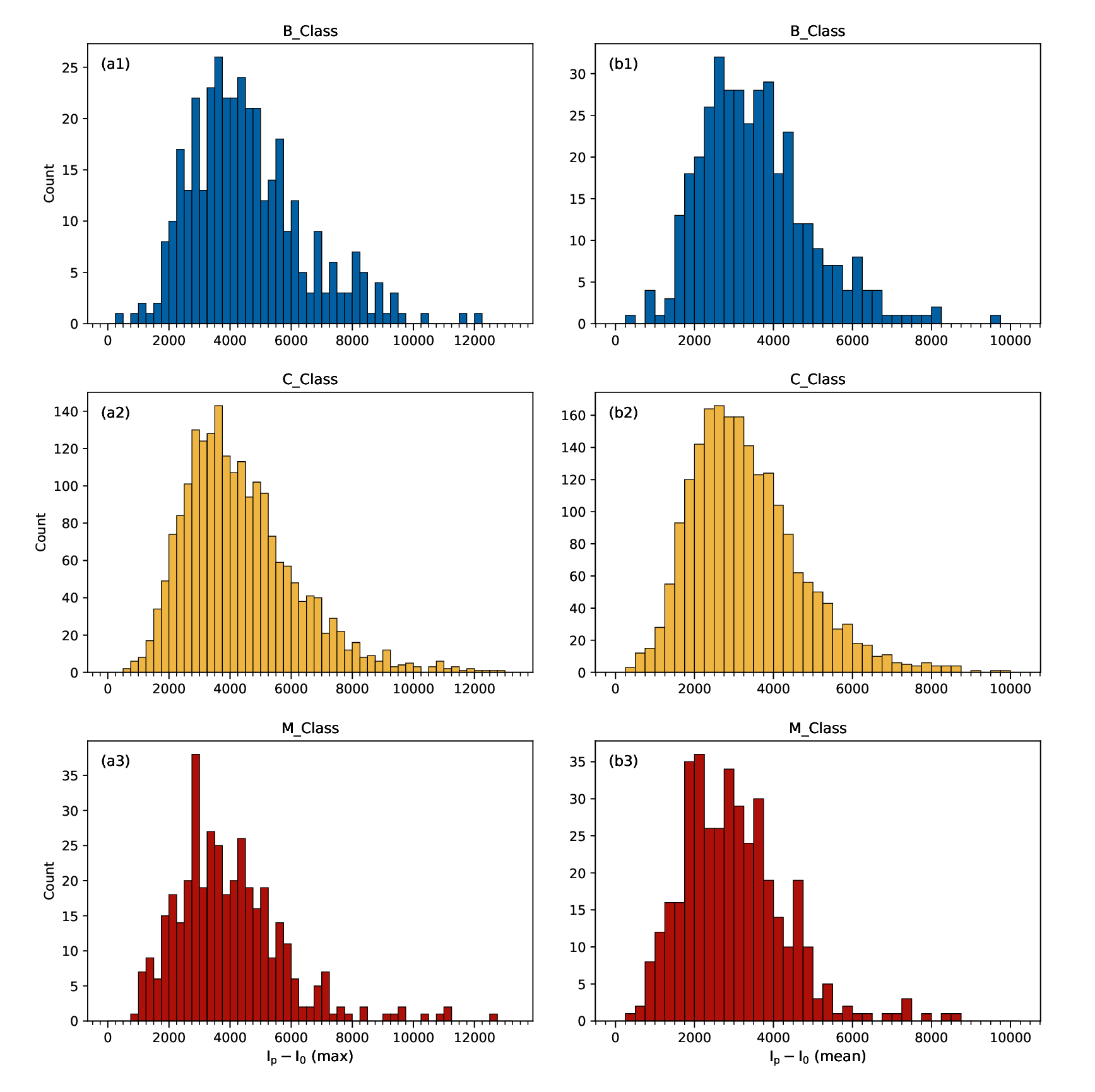}
\caption{
Same as Figure~\ref{fig10}, but for the absolute WL enhancements ($I_p-I_0$) in WL kernels for B-, C-, and M-class flares.
}\label{fig11}
\end{figure}

Figure \ref{fig4} presents representative examples of WLFs with different GOES classes, including three B-class, three C-class, and three M-class events. These WLFs contain varying numbers of WL kernels  located in different regions (umbra, penumbra, or quiet-Sun regions),  and these kernels exhibit distinctly different evolutionary behaviors. For the single-kernel cases, comparison with the GOES SXR flux shows that some WL kernels appear during the rising phase of the flare (panel (b1)), while others appear during the decay phase (panels (a1) and (a3)). For WLFs with two WL kernels, the two kernels may evolve synchronously, with their maxima coinciding with the SXR flux peak (panels (a2) and (b2)); alternatively, they can evolve asynchronously and peak at different stages of the flare (panel (c2)). For the multi-kernel cases, the evolutionary behavior becomes more complex: some kernels may evolve synchronously, such as K1 and K2 in the M3.3 flare (panel (c3)), while in other cases all kernels may be asynchronous (panels (a3) and (b3)). These observational results indicate that the continuum emission of WLFs exhibits rich and diverse evolutionary characteristics, and the temporal differences among WL kernels suggest that they correspond to distinct energy release processes within the flare.

Figure \ref{fig5} shows the occurrence of WLFs across different flare classes, as well as the fraction of ARs containing WLFs as a function of their flare numbers. In the panel (a), we see that the flares are mainly concentrated in the C1.0--C4.0 range, with the largest number of flares occurring around C1.0. As the flare class increases further, the number of flares generally decreases. M-class flares are mainly concentrated around M1.0, while higher-class flares are relatively rare. On this basis, the relative proportions of WLFs and non-WLFs also show clear variations among different flare-class bins. Panel (b) shows the WLF occurrence rate in different flare-class bins. To reduce the influence of small-number statistical fluctuations, the occurrence rate was calculated only for bins containing more than 10 flares. The results show that the WLF occurrence rate exhibits an overall increasing trend with flare class, although the variation is not strictly monotonic. In the B-class range, the WLF occurrence rate is generally low, mostly below 50\%. After entering the C-class range, the occurrence rate gradually increases, reaching about 60\%--70\% in the middle-to-high C-class bins. In the low M-class range, the WLF occurrence rate further increases to above 80\%. This refined statistical result further indicates that, in our sample, flares with higher magnitudes are more likely to produce detectable WL emission.

Panels (c) and (d) present the distribution of ARs with and without WLFs as a function of the number of flares they produced. We see the ARs without WLFs are mainly concentrated in the low-flare-count bins, particularly those producing only 1-2 flares. As the number of flares associated with an AR increases, the fraction of ARs hosting WLFs increases systematically: for ARs with 1, 2, 3, and 4 flares, the fractions are 43.6\%, 68.1\%, 76.5\%, and 86.1\%, respectively, and when the flare number reaches 6 or more, nearly 100\% of these ARs produce WLFs. It should be noted that the sample analyzed in this study consists entirely of low-activity ARs with relatively simple magnetic configurations, where the flares produced are predominantly of low to intermediate class. B- and C-class flares account for about 95.5\% of the sample, and the vast majority of them are below C5.0. Even in such weakly active environments, a flare number of 2 or more is sufficient for an AR to have a very high probability of hosting WLFs. It can therefore be inferred that ARs with more complex magnetic structures and stronger activity are even more likely to produce WLFs. This further demonstrates that WLFs are a very common phenomenon on the Sun.

Figure \ref{fig6} presents the distribution of WLFs with different numbers of WL kernels. From panel (a), we see the number of WLFs decreases rapidly as the number of their WL kernels increases.  An exponential fit to the WLFs distribution yields a goodness of fit of $R^2=0.9953$, indicating an approximately exponential decay. The fraction of WLFs containing a single WL kernel is 40.7\%, that containing two WL kernels is 22.8\%, and when the number of WL kernels reaches three, the corresponding fraction is only 11.6\%. These three categories together constitute 75.1\% of all WLFs, indicating that the majority of WLFs possess only a small number of WL kernels. When classified into single-kernel, double-kernel, and multi-kernel cases, their respective fractions are 40.7\%, 22.8\%, and 36.5\%. 

Panel (b) further compares the relative fractions of B-, C-, and M-class WLFs with different numbers of WL kernels. These distributions differ markedly among the three classes. For B-class WLFs, the fraction with a single WL kernel is the highest, approaching 60\%, and the number of WLFs decreases rapidly as the WL kernel number increases. For C-class WLFs, the fraction with a single WL kernel drops to about 40\%, and the corresponding decay trend becomes more gradual. For M-class WLFs, the fraction with a single WL kernel is less than 20\%, and the fractions change little from one to four kernels, with a rapid decrease occurring only when the WL kernel number exceeds four. These results indicate that higher-class WLFs tend to host multiple WL kernels, corresponding to more complex WL emission processes.

Similarly, figure \ref{fig7} shows the distribution of WLFs with different WL areas. We see the number of WLFs decreases rapidly as the WL emission area increases, and a power-law fit yields a goodness of fit of $R^2=0.9782$ (panel (a)). Comparing different classes, B-class WLFs have the highest fraction in the smaller WL emission areas, followed by C-class WLFs, with M-class WLFs being the lowest (panel (b)). This indicates that, as the flare class increases, WLFs tend to have larger WL areas.

Figure \ref{fig8} gives the statistics of WLFs and non-WLFs, and of the properties of WL kernels among B-, C- and M-class flares. Overall, the WLF occurrence rate increases significantly with flare class. The fraction of WLFs is highest for M-class flares (85.2\%), followed by C-class flares (55.9\%), and lowest for B-class flares (36.9\%), indicating that the WLF occurrence rate is closely related to total flare energy (panel (a)). In terms of the distribution of WL kernel numbers, B-class flares are dominated by single-kernel WLFs (56.9\%), whereas M-class flares are dominated by multi-kernel ($>3$) WLFs (69.3\%), suggesting that the number of WL kernels is tightly linked to the  energy budget of the parent flare (panel (b)). Regarding the evolutionary characteristics of double- and multi-kernel WLFs, asynchronous evolution is dominant across all classes, with corresponding fractions of 87.4\%, 90.5\%, and 98.4\% for B-, C-, and M-class flares, respectively (panel (c)). As for the spatial distribution of WL kernels, 91.4\% of WL kernels in B-class flares are located in the quiet Sun, 8.4\% in the penumbra, and only 0.2\% in the umbra; for C-class flares, the corresponding fractions are 86.1\%, 13.3\%, and 0.6\%; and for M-class flares, they are 74.4\%, 21.6\%, and 3.6\%. It can be seen that, as the flare class increases, WL kernels tend to shift from the quiet Sun toward the penumbra and umbra (panel (d)). Nevertheless, WL kernels remain predominantly located in the quiet Sun, which explains why WLFs have long been difficult to detect—most of their WL kernels lie in the quiet Sun, where the contrast with the background is weak and thus hard to identify directly by eyes.

Figure \ref{fig9} shows the distributions of the areas and durations of WL kernels for different classes of WLFs. From left panels, we see that the distributions of WL kernel areas for B-, C-, and M-class WLFs are generally similar. Most WL kernels have areas smaller than 2 ${\rm arcsec}^2$ (about 8 HMI pixels), and the maximum area does not exceed 5 ${\rm arcsec}^2$ (about 20 HMI pixels), indicating that these WL kernels are all small-scale structures. It should be noted that all areas have been corrected for projection effects. In terms of the number distribution of WL kernel areas, the number of WL kernels in C-class WLFs decreases rapidly with increasing area, but the decreasing trend becomes more gradual at about 1.9 ${\rm arcsec}^2$. The area distribution of WL kernels in M-class WLFs shows a similar trend, whereas the trend for B-class WLFs is less pronounced. This phenomenon may be due to the limited sample size, or it may indicate that additional heating mechanisms are involved for relatively large WL kernels. 

Regarding duration (right panels), the overall distributions of WL kernels for B-, C-, and M-class WLFs are also similar, approximately normal, with the bulk concentrated between 6 and 16 minutes and a peak at 10–12 minutes.  However, these distributions deviate from a perfect Gaussian shape. Taking C-class WLF kernels as an example, the right-hand side exhibits a slowly decaying extended tail at 17–22 minutes, and B- and M-class WLF kernels show similar trends. This suggests that longer-lived WL kernels may involve additional heating mechanisms.

Figure \ref{fig10} presents the statistical distribution of the relative WL enhancement of WL kernels in B-, C-, and M-class flares. For the maximum relative WL enhancement (left panels), the overall distributions of WL kernels for B-, C-, and M-class WLFs are similar: the maximum relative enhancement is mainly concentrated between 0.03 and 0.2, with a peak at 0.06–0.09.  It is worth noting that the enhancement thresholds adopted in many previous studies are typically around 0.08 \citep[e.g.][]{Hudson2006, Kuhar2016, Song2018b}, which coincides with the dominant enhancement range for most WL kernels in our sample. The distribution is approximately normal overall, but with slight deviations. Taking the WL kernels of C-class WLFs as an example, a slowly decaying extended tail can be seen at the right-hand side between 0.15 and 0.22, a feature also shared by B- and M-class WLF kernels. For the mean relative WL enhancement (right panels), the overall distributions of WL kernels for B-, C-, and M-class WLFs are likewise similar: the mean values are mainly concentrated between 0.03 and 0.15, with a peak at 0.06–0.08; the distribution is also approximately normal overall but slightly deviates, with the decline in the number of WL kernels becoming more gradual on the right-hand side between 0.11 and 0.19. This suggests that WL kernels with relatively large enhancements may involve additional heating mechanisms.

\begin{table*}[htbp]
\centering
\caption{Statistics of the relationships between the peak times of WL kernels and soft X-ray fluxes.}
\label{tab3}
{\fontsize{4.8}{8.8}\selectfont
\begin{tabular}{ccccccccccc}
\hline
\hline
\multirow{2}{*}{Flare class} & \multicolumn{3}{c}{$\Delta t = t_{\rm WL,peak} - t_{\rm dSXR,peak}$} & \multicolumn{3}{c}{$\Delta t^\prime = t_{\rm WL,peak} - t_{\rm SXR,peak}$}   \\
\cmidrule(lr){2-4} 
\cmidrule(lr){5-7} 
                                       & $\Delta t < -1\ min$ & $ -1\ min \le \Delta t \le 1\ min$ & $\Delta t > 1\ min$ & $\Delta t^\prime < -1\ min$ & $ -1\ min\le \Delta t^\prime \le 1\ min$ & $\Delta t^\prime > 1\ min$    \\
\hline
	B & 100 (19.6\%) & 126 (24.8\%) & 283 (55.6\%)   & 220 (43.2\%)  & 122 (24.0\%) & 167 (32.8\%)     \\
	C & 344 (18.3\%) & 446 (23.7\%) & 1091 (58.0\%) & 861 (45.8\%) & 470 (25.0\%) & 550 (29.2\%)  \\
	M & 52 (16.0\%)  & 93 (28.5\%) & 181 (55.5\%)    & 168 (51.5\%)  & 73 (22.4\%) & 85 (26.1\%)   \\
	Total   & 496 (18.3\%) & 665 (24.5\%) & 1555 (57.3\%) & 1249 (46.0\%) & 665 (24.5\%) & 802 (29.5\%)  \\
											
\hline
\end{tabular}
}
\begin{tablenotes}
        \tiny
        \item[1] Here $t_{\rm WL,peak}$ refers to the emission peak times of WL kernels. $ t_{\rm SXR,peak}$ and $t_{\rm dSXR,peak}$ denote the peak times of the GOES SXR flux and its time derivative, respectively.
      \end{tablenotes}
\end{table*}

\begin{figure}[h]
\centering
\includegraphics[width=0.9\textwidth]{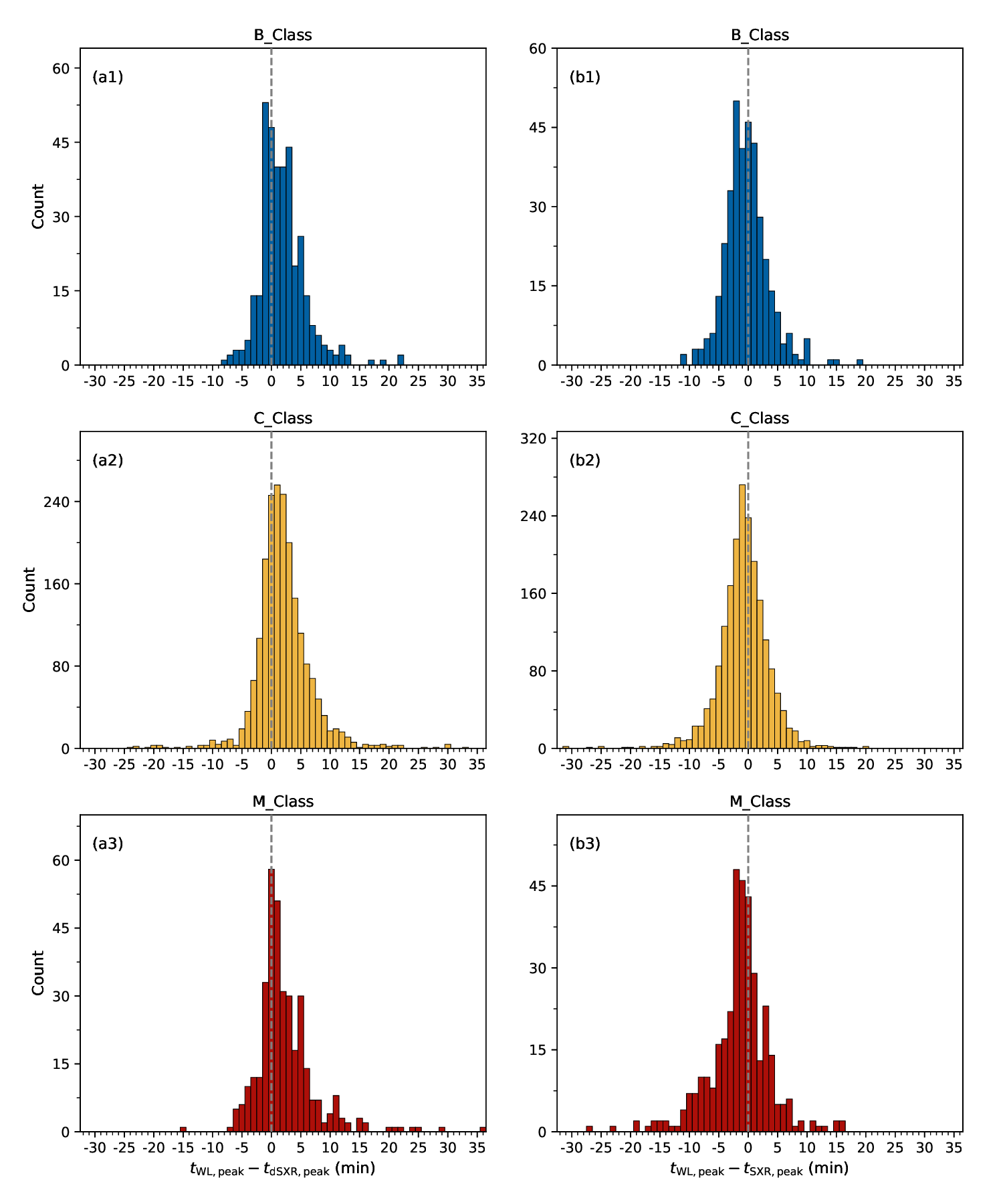}
\caption{
Statistics of the time differences between the peak times of WL kernels and GOES soft X-ray (SXR) fluxes for B-, C-, and M-class WLFs. Panels (a1)–(a3) show the time differences between the WL peaks ($t_{\rm WL,peak}$​) and the peaks of the GOES SXR flux derivative ($t_{\rm dSXR,peak}$​), while panels (b1)–(b3) show the time differences between the WL peaks and the peaks of the GOES SXR flux ($t_{\rm WL,peak}$​). The gray dashed line marks zero time difference.
}\label{fig12}
\end{figure}

Figure \ref{fig11} further shows the statistical distribution of the absolute WL enhancement of WL kernels in B-, C-, and M-class flares. Similar to the relative WL enhancement, the overall distributions of the absolute enhancement of WL kernels for B-, C-, and M-class WLFs are also approximately normal, but with slight deviations. For the maximum absolute WL enhancement (left panels), the values are mainly concentrated between 2000 and 8000 DN, with a peak at 3000–5000 DN. As for the mean absolute WL enhancement (right panels), the values are mainly concentrated between 1000 and 6000 DN, with a peak at 2000–4000 DN.

Figure \ref{fig12} shows the distributions of the time differences between the emission peaks of WL kernels and the GOES SXR fluxes for B-, C-, and M-class WLFs. From the comparison between the emission peak times of WL kernels and the peak times of the GOES SXR derivative flux (left panels), the overall distributions for B-, C-, and M-class WLFs are similar. The time differences ($\Delta t = t_{\rm WL,peak} - t_{\rm dSXR,peak}$) are mainly concentrated between -5 and 10 minutes, with peaks at -1 to 3 minutes, indicating that most WL kernel emission peaks lag behind the SXR derivative peak. Table \ref{tab3} lists the corresponding fractions of $\Delta t $. Given that the HMI data cadence is 45 s, we define $ -1\ min \le \Delta t \le 1\ min$ as synchronous, $\Delta t < -1\ min$ as leading, and  $\Delta t > 1\ min$ as lagging. Overall, synchronous, leading, and lagging cases account for 24.5\%, 18.3\%, and 57.3\%, respectively, and these fractions differ little among B-, C-, and M-class flares.

From the comparison between the WL kernel peak times and the GOES SXR peak times (right panels), the overall distributions are likewise similar across all classes. The time differences ($\Delta t^\prime = t_{\rm WL,peak} - t_{\rm SXR,peak}$) are mainly concentrated between -10 and 8 minutes, with peaks at -3 to 0 minutes, indicating that most WL kernel emission peaks lead the SXR flux peaks. The corresponding fractions in Table \ref{tab3} show that, overall, synchronous, leading, and lagging cases account for 24.5\%, 46\%, and 29.5\%, respectively. By class, M-class WLFs have the highest fraction of leading WL kernel peaks (51.5\%), followed by C-class (45.8\%) and B-class (43.2\%); the lagging fraction is highest for B-class WLFs (32.8\%), compared with 29.2\% for C-class and 26.1\% for M-class; the synchronous fractions are relatively similar across all classes.

If the Neupert effect \citep{Neupert1968} holds, i.e., the time derivative of the SXR flux corresponds to the variation in hard X-ray (HXR) emission, the above results indicate that about half of the WL kernel emission peaks lie between the HXR and SXR peak times, while roughly 25\% each lead the HXR peak or lag the SXR peak.

\begin{figure}[h]
\centering
\includegraphics[width=0.9\textwidth]{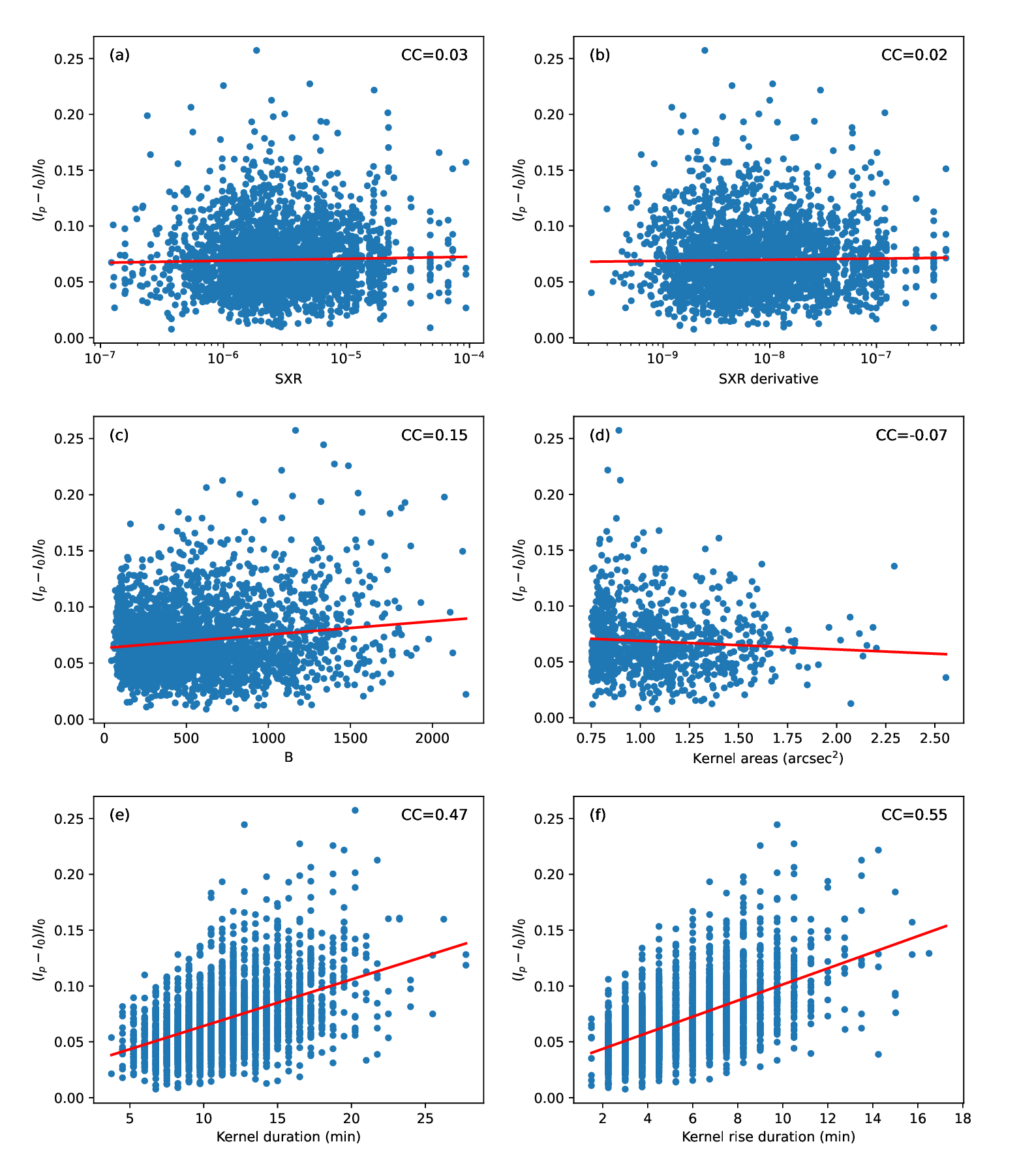}
\caption{
Scatter plots of the relative WL enhancement ($(I_p-I_0)/I_0$​) of WL kernels versus (a) the GOES SXR peak flux, (b) the GOES SXR flux derivative peak, (c) the magnetic field strength (B), (d) the WL kernel area, (e) the WL kernel duration, and (f) the WL rising duration.
}\label{fig13}
\end{figure}

Figure \ref{fig13} presents the relationships between the relative WL enhancement of WL kernels and the following parameters: the SXR peak flux and the peak of its time derivative, the magnetic field strength at the WL kernel, the WL kernel area, the WL duration, and its rising period. The results show that the WL kernel intensity exhibits no significant correlation with the SXR peak flux or the peak of its time derivative, the local magnetic field strength, or the kernel area (panels (a)-(d)). It indicates that the radiative intensity of WL kernels bears no dependence on flare class, local magnetic field strength, or the spatial scale of WL kernels. In contrast, the WL kernel intensity shows a clear correlation with its duration and rise time (panels (e) and (f)), suggesting that the WL kernel intensity is closely related to the duration and efficiency of heating.

\section{Summary and discussion}\label{sec5}

Whether WLFs are a common phenomenon on the Sun has long been debated in the solar physics community \citep[e.g.][]{Neidig1989, Ding2007}. In the past decades, WLFs were very rare in observations and were long considered to represent extreme eruptive events \citep[e.g.][]{McIntosh1972, Neidig1983, Fang2013}. However, it has also been pointed out that the properties of telescope, such as its sensitivity, spatial resolution, and temporal resolution, can also limit the detection of WLFs \citep[e.g.][]{Jess2008}. The detection methods for WLFs are also of great importance. Recently, several new methods have been proposed \citep[e.g.][]{Mravcova2017, Cai2024, Sharma2026}. However, these methods suffer from various limitations—for example, they can only detect large WLFs, require different thresholds for different flare classes, involve complex mathematical procedures, or depend on artificial intelligence.

To address this question, we first developed a physically motivated detection method based on WL emission pulses. The method is simple and broadly applicable, using a uniform threshold across different flare classes without complex mathematical operations or artificial intelligence \citep{Lishuanghong2026}. Together with the HMI's nearly 15 years of stable, continuous photospheric observations, which essentially cover most of Solar Cycle 24 and a large part of Solar Cycle 25, it provides us an excellent opportunity to answer this question. Moreover, our flare sample is drawn from low-activity ARs (flare number $<$ 10), which represent the lowest level of solar activity, with flares mainly of B- and C-class, a small number of M-class, and no X-class flares. This differs from many previous WLF studies, which mainly focused on M- and X-class flares from highly complex ARs \citep[e.g.][]{Kuhar2016, Namekata2017, Cai2024, Liying2024b, Jing2024, Jing2025}. Therefore, investigating flares in these ARs will better reflect the ubiquity of WLFs on the Sun, enabling us to answer this long-standing question more reliably than ever before.

A total of 1951 flares (547 B-class, 1316 C-class, and 88 M-class) from 769 low-activity ARs during January 2011 – May 2025 were investigated. Of these, 1012 flares were identified as WLFs, corresponding to an overall occurrence rate of 51.9\%. The WLF occurrence rate increases clearly with flare class, reaching 36.9\%, 56.9\%, and 85.2\% for B-, C-, and M-class flares, respectively. For the first time, we find that WL emission is widespread in B-class flares. Previous reports of WLFs were mainly concentrated on M- and X-class flares, with only a few involving C-class flares. Our investigation shows that WLFs are not only widespread among C-class flares but also common among B-class flares, indicating that WL emission is not exclusive to large flares. The increase in WLF occurrence rate with flare class (Fig.\ref{fig5} (a) and (b)) suggests that the total flare energy plays an important role in the occurrence of WLFs.

Among these low-activity ARs, 508 produced WLFs, accounting for 66.1\%. According to our statistics, when an AR contains three flares, its probability of producing a WLF reaches $\sim$80\% (Fig.\ref{fig5} (c) and (d)). This further suggests that the vast majority of ARs are capable of producing WLFs. The WLF occurrence rate exhibits clear year-to-year variation and shows a strong positive correlation with the annual mean sunspot number of the selected ARs, with a correlation coefficient of $r\approx0.85$ (Fig.\ref{fig2}). This indicates that the occurrence of WLFs is closely related to the complexity of the AR magnetic field.

We conducted a systematic statistical comparison of the WL kernels in these WLFs, including their number, area, duration, WL enhancement amplitude, and time differences relative to the peaks of the GOES SXR flux and its derivative. In terms of WL kernel number, B-class WLFs are dominated by single-kernel events, with the highest fraction (57\%), followed by C-class (39\%) and M-class (17\%). Moreover, the corresponding fractions decrease rapidly as the WL kernel number increases (Fig.\ref{fig6} and Fig.\ref{fig8}(b)). This indicates that higher-class flares tend to possess more WL kernels, implying that flares with greater energy are more likely to heat the lower atmosphere at multiple locations. For WLFs with two or more WL kernels, the evolution of these kernels is generally asynchronous (Fig.\ref{fig8}(c)), indicating temporal differences in lower-atmosphere heating among different regions. Regarding kernel location, we find that WL kernels are predominantly located in the quiet Sun (Fig.\ref{fig8}(d)), which explains why WLFs have long been difficult to detect—the contrast of WL emission in these regions is very weak, making it hard to identify either by eye or with conventional threshold methods.

The overall distributions of the area, duration, and enhancement amplitude of WL kernels show no significant differences among B-, C-, and M-class flares (Fig.\ref{fig9}, \ref{fig10} and \ref{fig11}). This result implies that WL emission is more likely to be a localized heating phenomenon. The total flare energy not significantly affecting  the intrinsic properties of individual WL kernels. It should be noted, however, that our sample is dominated by relatively weak flares (B- and C-class). If more M- and X-class flares are included, the properties of WL kernels may differ significantly. Regarding the relationship between the WL kernel emission peaks and the peaks of the SXR flux and its derivative, we find that about half of the WL kernel peaks lie between the SXR derivative peak and the SXR peak (Fig.\ref{fig12} and Table \ref{tab3}), suggesting that WL emission is more likely a consequence of coupled nonthermal and thermal processes. It should also be noted, however, that the GOES SXR flux is a globally integrated quantity, whereas the evolution of WL kernels is a localized phenomenon, and a direct comparison between them may thus carries inherent limitations.

Considering that the HMI data have a cadence of 45 s and a spatial resolution of $\sim$$0.5^{\prime\prime}$, we cannot identify WL emission with shorter durations (< 45 s) or smaller spatial scales (< $0.5^{\prime\prime}$). In addition, to exclude random heating events, we discarded isolated WL points ($\le 2$ pixels), which also artificially removes WL emission on smaller scales. Given these limitations, and based on our investigation of WL emissions in 1951 flares (547 B-class, 1316 C-class, and 88 M-class)  from 769 low-activity ARs observed during January 2011 – May 2025, we strongly conclude that WLFs are a very common phenomenon on the Sun. It further indicates that stellar flares may be more prevalent than previously recognized. Many very weak stellar flares are difficult to discern in integrated WL spectra, and stars with no detected flares may nonetheless host flare activity comparable to that of the Sun, yet such flares are hard to observe in the WL band. This underscores the necessity of multi-wavelength coordinated observations in the search for stellar flares.

A more noteworthy question is why the WL emission produced by different flares differs markedly in intensity and spatial scale. These difference seem not directly related to flare class or total flare energy (Fig.\ref{fig13}). This implies that other factors modulate the heating of the lower atmosphere during flares, among which the most influential is likely the magnetic topology at the time of the flare. Magnetic configuration strongly affects the height, spatial scale, heating efficiency, and energy transport pathways of the flare energy, thereby leading to different responses in the lower atmosphere. Future dedicated studies of these WL kernels will further advance our understanding of the thermodynamic processes in the lower solar atmosphere.


\backmatter

\bmhead{Acknowledgements}

This work is supported by National Key R\&D Program of China (Grant No. 2022YFF0503001, 2021YFA0718601), National Natural Science Foundation of China (Grant No. 12173049, 12473053, 12373057). K.W acknowledges support from the National Natural Science Foundation of China (Grant No. 12373035) and the Sichuan Science and Technology Program (Grant No. 2026YFTX0025). The authors thank the SDO and GOES teams for providing the data. SDO is a space mission in the Living With a Star Program of NASA.


\bibliography{sn-bibliography}

\end{document}